\documentclass[journal]{IEEEtran}
\usepackage{multirow}
\usepackage{booktabs}
\usepackage{amsfonts}
\usepackage{amsmath}
\usepackage{amsmath,cases}
\usepackage{amssymb}
\usepackage{graphicx}
\usepackage{cite}
\usepackage{epsfig}
\usepackage{url}
\usepackage{algorithm}
\usepackage{algorithmicx, algpseudocode}
\usepackage{epstopdf}
\usepackage{color}
\usepackage{mathrsfs}
\usepackage[justification=centering]{caption}
\usepackage{stfloats}
\usepackage{diagbox}
\usepackage{subcaption}
\usepackage{balance}

\newcommand{\ds}{\displaystyle}

\newcommand{\la}{\langle}
\newcommand{\ra}{\rangle}

\newcommand{\bX}{\boldsymbol{X}}
\newcommand{\bx}{\boldsymbol{x}}

\newcommand{\clC}{{\cal C}}

\newcommand{\clN}{{\cal N}}

\newcommand{\clK}{{\cal K}}

\newcommand{\balpha}{\pmb{\alpha}}

\newcommand{\clD}{{\cal D}}

\newcommand{\bz}{\mathbf{z}}

\newcommand{\bY}{\mathbf{Y}}

\newcommand{\xio}{x^{(\iota+1)}}

\newcommand{\Xii}{X^{(\iota)}}
\newcommand{\Xio}{X^{(\iota+1)}}
\newcommand{\bV}{\mathbf{V}}

\newcommand{\bW}{\mathbf{W}}

\newcommand{\Pii}{\Pi^{(\iota)}}
\newcommand{\ri}{r^{(\iota)}}
\newcommand{\ai}{a^{(\iota)}}

\newcommand{\Vi}{V^{(\iota)}}

\newcommand{\bi}{b^{(\iota)}}

\newcommand{\tri}{\tilde{r}^{(\iota)}}

\newcommand{\chii}{\chi^{(\iota)}}
\newcommand{\chio}{\chi^{(\iota+1)}}

\newcommand{\fii}{f^{(\iota)}}

\newcommand{\varphii}{\varphi^{(\iota)}}

\newcommand{\tvarphii}{\tilde{\varphi}^{(\iota)}}

\newcommand{\lambdai}{\lambda^{(\iota)}}

\newcommand{\tfi}{\tilde{f}^{(\iota)}}

\newcommand{\clAio}{\mathcal{A}^{(\iota+1)}}

\newcommand{\clM}{\mathcal{M}}
\newcommand{\Wi}{W^{(\iota)}}
\newcommand{\Wio}{W^{(\iota+1)}}
\newcommand{\Yi}{Y^{(\iota)}}
\newcommand{\clCi}{\clC^{(\iota)}}
\newcommand{\bchi}{\pmb{\chi}}
\newcommand{\chiio}{\chi^{(\iota+1)}}
\newcommand{\clHi}{\mathcal{H}^{(\iota)}}
\newcommand{\tclCi}{\tilde{\mathcal{C}}^{(\iota)}}

\newcommand{\clBi}{\mathcal{B}^{(\iota)}}
\newcommand{\tclBi}{\tilde{\mathcal{B}}^{(\iota)}}
\newcommand{\tclDi}{\tilde{\mathcal{D}}^{(\iota)}}
\newcommand{\tfii}{\tilde{f}^{(\iota)}}
\newcommand{\clG}{\mathcal{G}}
\newcommand{\clGi}{\mathcal{G}^{(\iota)}}
\newcommand{\mI}{\mathbb{I}}
\allowdisplaybreaks
\begin{document}
\title{Holographic Multi-User Multi-Stream Beamforming Maintaining Rate-Fairness}
\author{W. Zhu$^{1}$, H. D. Tuan$^1$, E. Dutkiewicz$^1$,  H. V. Poor$^2$, and L. Hanzo$^3$
\thanks{The work was supported in part by the Australian Research Council's Discovery Projects under Grant DP190102501,  in part by the U.S National Science Foundation under Grants CNS-2128448 and ECCS-2335876, in part by  the Engineering and Physical Sciences Research Council projects EP/W016605/1, EP/X01228X/1 and EP/Y026721/1 as well as of the European Research Council's Advanced Fellow Grant QuantCom (Grant No. 789028).}
\thanks{$^1$School of Electrical and Data Engineering, University of Technology Sydney, Broadway, NSW 2007, Australia (email: wenbo.zhu@student.uts.edu.au, tuan.hoang@uts.edu.au, eryk.dutkiewicz@uts.edu.au); $^2$Department of Electrical and Computer Engineering, Princeton University, Princeton, NJ 08544, USA (email: poor@princeton.edu);
$^3$School of Electronics and Computer Science, University of Southampton, Southampton, SO17 1BJ, U.K (email: lh@ecs.soton.ac.uk) }
}
\date{}
\maketitle
{\color{black}
\begin{abstract}
We present the first investigation into the transmission of multi-stream information from a base station equipped with reconfigurable holographic surfaces (RHS) to multiple users with the aid of multi-antenna arrays. Building upon this, we propose the joint design of RHS and baseband beamformers that enables multi-stream delivery at fair rates across all users. Specifically, we first introduce a max-min rate optimization approach, which aims for maximizing the minimum rate for all users through iterative solutions of quadratic problems. To reduce complexity, we then propose a surrogate-based optimization approach that offers a low-complexity design alternative relying on closed-form updates. Our simulations show that the surrogate-based approach achieves nearly the same minimum rate as max-min optimization, while delivering sum-rates comparable to those of sum-rate maximization, overcoming the rate-fairness deficiency typical of the latter.
\end{abstract}
}

\section{Introduction}
In recent years, the field of wireless communications and sensing has witnessed a significant transformation with the advent of metasurfaces harnessed as  massive antenna arrays \cite{Panetal21,Lonetal21,DD21,Denetal21wc,Zhaetal22twc, Zhaetal23,Denetal23tvtmag,ZZDS23,Denetal23jsac}. Beamforming techniques conceived for creating directional beams for specific end-users are capable of supporting a massive number of connections and services in the Internet-of-Things. However, the beamformers relying on reconfigurable intelligent surfaces (RISs) \cite{Panetal21,Lonetal21,Zhaetal23} and on dynamic metasurfaces \cite{DD21,Zhaetal22twc, Zhaetal23} employing phased arrays, known for their high power consumption and excessive manufacturing costs. By contrast, beamforming relying on reconfigurable holographic  surfaces (RHSs) utilizes  low-cost electronically scanned arrays for manipulating the radiation amplitudes \cite{P12,Johetal15,Denetal21wc,Denetal23tvtmag,Denetal23jsac}. However, holographic beamforming design relies on a considerable number of optimization variables and constraints involved in the radiation amplitude optimization.  For instance, a holographic beamformer having a moderate-sized RHS of  $12\times 12$ elements and eight feeds already entails $12^2\times 8=1152$ optimization variables subject to the corresponding amplitude constraints. Consequently, holographic beamforming leads to a computationally demanding large-scale nonconvex optimization problem that continues to challenge existing state-of-the-art algorithms, motivating the need for scalable and efficient solutions.
It is fair to say that holographic beamforming  is still in its early stages of development \cite{D21,Denetal21tvt,Huetal22,Denetal22twc,Denetal23tvtmag,WCMW24}.
The pioneering paper \cite{Denetal22twc} utilized fractional programming for the alternating optimization of a  holographic beamformer using the sum-rate objective function. However, optimizing this fractional programming-based convex problem does not necessarily result in an improved sum-rate, potentially leading to convergence issues in the alternating optimization procedure iterating between the baseband (digital) beamformer and the holographic beamformer. In the simulation scenarios of \cite{Denetal22twc}, a single data stream was transmitted to
three $10$-antenna users, where at least four feeds were needed to accommodate the baseband zero-forcing beamforming where the size of the RHS ranged from $144$ to $1600$. In this convex formulation, the number of optimization variables and constraints was between $144\times 4=576$ and $1600\times 4=6400$, resulting in a prohibitive   complexity. It should be emphasized that the sum-rate problem has frequently been considered in multi-user communications due to its relatively low computational complexity. Specifically, the smoothness of the the sum-rate function enables efficient beamforming design through iterative evaluation of low-complexity closed-form solutions. Under the same allocation of resources (such as transmit power or bandwidth), the maximized sum-rate continues to increase with the number of users. However, this approach leads to near-zero rates for certain users, which is primarily because of prioritizing specific users having high rates owing to their high channel quality, while  sacrificing  others by assigning close to zero rates.  Consequently, sum-rate maximization sacrifices rate-fairness amongst users. Against this background, we lay down the computational foundations of holographic beamforming designs conceived for serving multiple users while maintaining rate-fairness. More explicitly, we have the following contributions:
\begin{itemize}
\item We tackle the challenge of maximizing the minimum rate among users (max-min rate optimization) to improve their rate-fairness. In contrast to \cite{Denetal22twc}, which considered scenarios involving the transmission of single information stream to multi-antenna users, our approach allows for multiple-stream transmissions. The user rate is now defined by the logarithmic determinant (log-det) of a nonlinear matrix-valued function of both the holographic and the baseband beamformers. Fractional programming is unsuitable for their optimization. In this context, we propose an alternating optimization algorithm harnessing convex quadratic programming for iteratively generating gradually improved feasible baseband and holographic beamformers, while ensuring algorithmic convergence.
\item We introduce a new penalized optimization framework that incorporates amplitude constraints in holographic beamformers into the optimization objective function. This reformulation tends itself to a low-complexity alternating optimization solution that maximizes the sum-rate by iteratively evaluating low-complexity closed-form solutions.
\item To circumvent the cubically escalating complexity of convex quadratic solvers harnessed for addressing the non-smooth minimum rate function, we unveil a new surrogate optimization objective and develop a computational method based on iterating by evaluating new closed-form expressions, hence resulting in low computational complexity. Optimizing this surrogate objective results in high minimum user-rates and also achieves sum-rates that are higher than those of conventional approaches. Hence, our new surrogate optimization strategy meets three key objectives: providing improved rate-fairness relative to existing methods, maximizing the sum-rate, and offering lower computational complexity than traditional solvers. Hence, the proposed surrogate optimization strategy is particularly suitable for beamforming in multi-user networks of realistic sizes, where it achieves high minimum user-rates, improved fairness, and low computational complexity.
\end{itemize}
The paper is organized as follows. Section II presents the system model and analyses the user rates achieved through baseband and holographic beamforming. Section III focuses on designing holographic and baseband beamformers for multiple multi-antenna users by maximizing their minimum rate. This section develops an alternating optimization procedure based on convex quadratic solvers to generate a sequence of improved holographic and baseband beamformers, while guaranteeing convergence.
Section IV introduces a new penalized optimization procedure that iterates by evaluating closed-form expressions for maximizing the sum-rate. In Section V, a new surrogate optimization problem is conceived for attaining both high minimum rate and sum-rate, which may not be achievable by max-min rate or by sum-rate maximization alone. Importantly, the optimization of this objective function is achieved by devising an efficient computational procedure, relying on iterating closed-form expressions of scalable complexity to find the optimal solution. This procedure is eminently suitable for the design of holographic and baseband beamformers of any scale.
Section VI provides simulations for verifying the accuracy of our results presented in the previous sections, while Section VII concludes the paper. Appendices are provided for the basic inequalities used in deriving the results of Sections III-IV and for lengthy derivations.

\emph{Notation.\;} Boldface fonts are specifically reserved for optimization variables.\footnote{The boldface font used for optimization variables explicitly distinguishes them from feasible points in the algorithm design. This enhances the prominence of optimization variables and helps to identify the specific structures such as quadratic forms in the associated functions. These variables may represent scalars, vectors, or matrices.}
We have $\la X,Y\ra\triangleq {\sf trace}(X^HY)$ for the matrices $X$ and $Y$. We also use $\la X\ra$ for the trace of $X$ when $X$ is a square matrix. $X\succ 0$ ($X\succeq 0$, resp.) means that $X$ is a Hermitian symmetric positive definite (positive semi-definite, resp.) matrix. As such $X\prec Y$ means that we have $Y-X\succ 0$.
$||X||$ is the Frobenius norm of the matrix $X$, which is defined by $\sqrt{\la XX^H\ra}$. Furthermore,
$[X]^2$ stands for $XX^H\succeq 0$, so $||X||^2=\la [X]^2\ra$. When $X\succeq 0$, its square root $\sqrt{X}$ is a positive semi-definite matrix satisfying $[\sqrt{X}]^2=X$. The symbol $|X|$ represents the determinant of the matrix $X$, so $\ln|X|$ denotes the natural logarithm of the determinant (log-det), while
${\sf vec}(X)$ stacks the  columns of the matrix $X$ into a single vector. The symbol
$\mI$ is commonly used to represent the identity matrix. However, we also use  $\mI_D$ to emphasize that its size is $D\times D$. Lastly, a summary of basic notations used throughout the paper along with their descriptions are presented in Table \ref{table:notation} for a quick reference.
\begin{table*}[!t]
\centering
\caption{Summary of Basic Notations}
\begin{tabular}{ll}
\toprule
Notation & Description \\
\midrule
$K\big/\clK$ & number of feeds on RHS $\big/$ index set $\{1,\dots, K\}$\\
$M\big/\clM$ & number of vertical and horizontal elements on RHS $\big/$ index set $\{1,\dots, M\}$\\
$N_u\big/\clN_u$ & number of users $\big/$ index set $\{1,\dots, N_u\}$\\
$D$ & number of each user' antennas \\
$H_{\nu}$ & channel spanning from the BS to user $\nu$\\
$\mu_{k,m,m'}$ & normalized amplitude of feed $k$ and $(m,m')$-th element on RHS\\
$\bx_{k,m,m'}$ & scaled amplitude of feed $k$ and $(m,m')$-th element on RHS\\
$\bX$ & holographic beamformer\\
$\bW_{\nu}$ & user $\nu$' baseband beamformer\\
$\bW\triangleq (\bW_1,\dots, \bW_{N_u})$ & set of baseband beamformer\\
$\bchi$ & auxiliary variable in penalized optimization\\
$r_{\nu}(\bW,\bX)$ & rate for user $\nu$\\
$\Wio$, $\Xio$, $\chio$ & values of $\bW$, $\bX$, $\bchi$ generated at $\iota$-th iteration\\
\bottomrule
\end{tabular}
\label{table:notation}
\end{table*}

\section{System Model}
We begin by recalling the holographic interference principle of \cite{Fonetal10}, which forms the basis of holographic beamforming.
\begin{figure}[t]
	\centering
	\includegraphics[width=8cm]{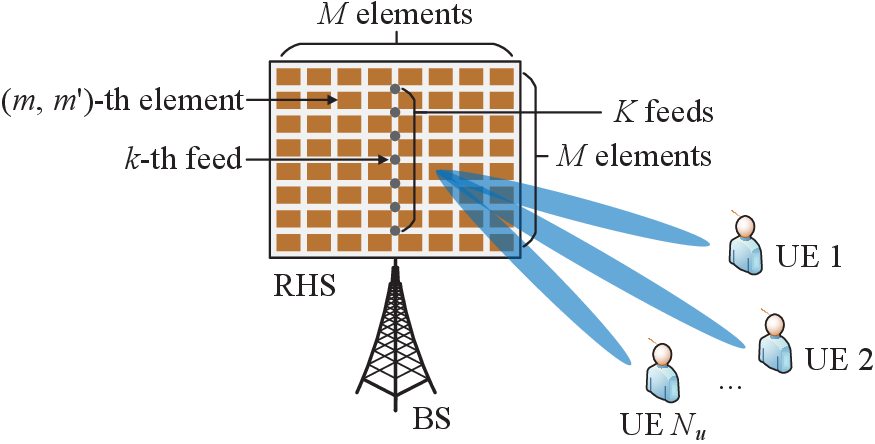}
	\caption{Illustration of RHS-assisted multi-user communication system.}
	\label{fig:RHS_model}
\end{figure}

Consider an RHS-assisted multi-user communication system, as depicted in Fig. \ref{fig:RHS_model}. An RHS consists of $K$ feeds indexed by $k\in\clK \triangleq \{1,\dots, K\}$ and a planar array of $M\times M$ elements, indexed by $(m,m')\in \clM\times \clM$ with $\clM\triangleq \{1,\dots, M\}$. The feeds are embedded in the bottom layer to generate incident electromagnetic waves, while the planar array is located at the Cartesian coordinates $(x,y,z)$, where  the RHS resides in the  $(x,y)$-plane. At the $(m,m')$-th radiation, the desired wave propagating in the direction $(\theta,\varphi)$ is represented by $\Phi(p_{m,m'},\theta,\varphi)=e^{-\jmath \la\kappa(\theta,\varphi), p_{m,m'}\ra}$, where $\kappa(\theta,\varphi)$ is the desired directional propagation vector in free space and $p_{m,m'}$ is the position vector of the $(m,m')$-th element. The reference wave generated by feed $k$ is
$\Phi_k(d^k_{m,m'})=e^{-\jmath \la\kappa_s,d^k_{m,m'}\ra}$, where $k_s$ is the propagation vector of the reference wave, and  $d^k_{m,m'}$ is the distance vector spanning from  feed $k$ to the $(m,m')$-th element.
The interference between the reference wave and the desired object wave is defined as
\begin{equation}\label{holo1}
\psi_{k,m,m'}\triangleq \Phi(p_{m,m'},\theta,\varphi)\Phi^*_k(d^k_{m,m'}).
\end{equation}
The holographic pattern is induced by the reference wave $\Phi_k(d^k_{m,m'})$ to engender
the wave propagation $\Phi(p_{m,m'},\theta,\varphi)|\Phi_k(d^k_{m,m'})|^2$. Accordingly, the radiation amplitude of the $(m,m')$-th element to generate the object wave in the radiation direction $(\theta,\varphi)$ is defined as $\Re\{\psi_{k,m,m'}\}=\cos\left(\la \kappa(\theta,\varphi),p_{m,m'}\ra-\la\kappa_s,d^k_{m,m'}\ra\right)$, which is then normalized as
\begin{equation}\label{holo2}
\mu_{k,m,m'}\triangleq \frac{\Re\{\psi_{k,m,m'}\}+1}{2}.
\end{equation}
The holographic beamforming effect is achieved by scaling the magnitude of $\mu_{k,m,m'}$ using
\begin{equation}\label{mag1}
\balpha_{k,m,m'}\in [0,1], (k,m,m')\in\clK\times\clM\times\clM.
\end{equation}
Accordingly, we define $\bx_{k,m,m'}=\balpha_{k,m,m'}\mu_{k,m,m'}$ and form
\begin{equation}\label{holo3}
\bx_k\triangleq {\sf vec}\left(\begin{bmatrix}\bx_{k,m,m'}\end{bmatrix}_{(m,m')\in\clM\times\clM}\right)\in\mathbb{R}^{M^2}, k\in\clK.
\end{equation}
Then
\begin{equation}\label{holo5}
\bX\triangleq \begin{bmatrix} \bx_1&\dots&\bx_K\end{bmatrix}\in\mathbb{R}^{M^2\times K}\ \&\
\bx\triangleq {\sf vec}(\bX)=\begin{bmatrix}\bx_1\cr
\dots\cr
\bx_K\end{bmatrix}
\end{equation}
represents the holographic beamformer, which is the objective of our holographic beamformer design.

Next, we use the above RHS at a base station (BS) to serve
$N_u$ users $\nu\in\clN_u\triangleq \{1,\dots, N_u\}$ in the downlink. Each of the users is
equipped with a $D$-antenna array, where $D\leq K$. Let $s_{\nu}\in\mathbb{C}^{D}$ along with $\mathbb{E}(s_{\nu} s^H_{\nu})=\mI_D$ be the multiple information stream intended for
user equipment (UE) $\nu$. Given the holographic beamformer $\bX$ defined in (\ref{holo5}),  the hybrid beamformer for
$s_{\nu}$ is in the form
\begin{equation}\label{holo6}
\bX\bW_{\nu}s_{\nu}
\end{equation}
with the baseband beamformer given by:
\begin{equation}\label{holo7}
\bW_{\nu}\triangleq [\bW_{\nu}(k,d)]_{(k,d)\in\clK\times \clD}\in \mathbb{C}^{K\times D}, \nu\in\clN_u.
\end{equation}
Let
\[
H_{\nu}=\begin{bmatrix}h_{\nu,1}\cr
\dots\cr
h_{\nu,D}  \end{bmatrix}
\in \mathbb{C}^{D\times M^2}; h_{\nu,d}\in\mathbb{C}^{1\times M^2}, d\in\clD\triangleq \{1,\dots, D\}
\] represent the channel spaning from the BS to user
$\nu\in\clN_u$, whose full channel state information  is assumed to be available.
Then the signal received at UE $\nu$ is formulated as:
\begin{equation}\label{holo8}
y_{\nu}=H_{\nu}\sum_{\nu'\in\clN_u}\bX\bW_{\nu'}s_{\nu'}+n_b,
\end{equation}
where $n_b$ is the background noise of power $\sigma$. Thus, for $\bW\triangleq (\bW_1,\dots, \bW_{N_u})$, the rate of $s_{\nu}$ (in nats) is defined by the log-det function:
\begin{equation}\label{holo9}
r_{\nu}(\bW,\bX)\triangleq \ln\left|\mI_D+[H_{\nu}\bX\bW_{\nu}]^2\clG^{-1}_{\nu}(\bW,\bX)\right|,
\end{equation}
where $\clG_{\nu}(\bW,\bX):\mathbb{C}^{K\times (DN_u)}\times \mathbb{R}^{M^2\times K}\rightarrow \mathbb{C}^{D\times D}$ is the so called interference covariance mapping \cite{Tametal17twc} defined by
\begin{equation}\label{icm}
\clG_{\nu}(\bW,\bX)\triangleq \sum_{\nu'\neq \nu} [H_{\nu}\bX\bW_{\nu'}]^2+\sigma \mI_D.
\end{equation}

\begin{figure}[t]
\centering
\includegraphics[width=6cm]{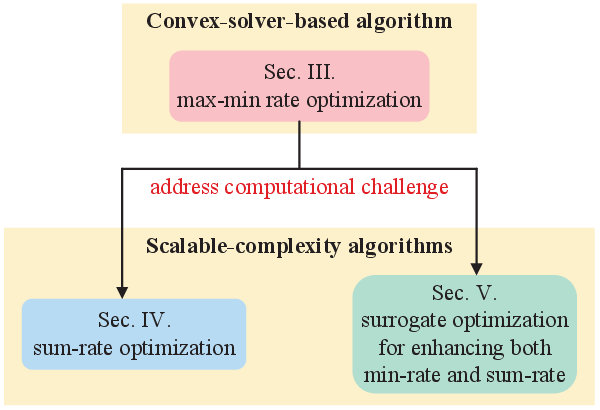}
\caption{Flowchart of the proposed algorithms.}
\label{fig:flowchart}
\end{figure}

In the given holographic multi-user multi-stream beamforming framework, we first investigate the max-min rate optimization problem to ensure fairness in users’ rates. To reduce computational complexity, we then tackle the sum-rate maximization problem using a penalized optimization approach for efficient computation. Additionally, we introduce a surrogate optimization formulation that efficiently enhances both the minimum rate and sum-rate. For clarity, Fig. \ref{fig:flowchart} presents a flowchart of the proposed algorithms.

\section{Max-min rate optimization of holographic beamforming}
Define the minimum rate function as
\begin{equation}\label{minrate}
f(\bW,\bX)\triangleq \min_{\nu\in\clN_u} r_{\nu}(\bW,\bX).
\end{equation}
We then address the following max-min rate optimization problem, which aims to maintain fairness across users' rates:
\begin{subequations}\label{holo10}
\begin{eqnarray}
\max_{\bW,\bX}\ f(\bW,\bX)\label{holo10a}\\
\mbox{s.t.}\quad \sum_{\nu\in\clN_u}||\bX\bW_{\nu}||^2\leq P, \label{holo10b}\\
\bx_{k,m,m'}\in [0,\mu_{k,m,m'}], (k,m,m')\in\clK\times\clM\times\clM,\label{holo10c}
\end{eqnarray}
\end{subequations}
where (\ref{holo10b}) enforces the power constraint within the given power budget $P$.
This problem is computationally demanding, not only because of the large number of optimization variables but also due to the intricate structure of the objective function. For instance, under a typical setting of $(K,M,D,N_u)=(4,12,2,4)$, the baseband beamformer involves $32$ complex variables, while the holographic beamformer requires $576$ real variables. Moreover, the objective function $f(\bW,\bX)$ in (\ref{holo10}) is not only highly nonlinear but also nonsmooth, as it represents the pointwise minimum of log-det functions.

We now proceed to develop a computational algorithm based on alternating optimization between the baseband and holographic beamformers, leveraging convex quadratic programming to address problem (\ref{holo10}).

Let $(W^{(0)}, X^{(0)})$ denote an initial feasible point of (\ref{holo10}), and suppose that $(W^{(\iota)}, X^{(\iota)})$ is the feasible point obtained at the $(\iota-1)$-st iteration. The alternating procedure then generates the next feasible point $(W^{(\iota+1)}, X^{(\iota+1)})$ at the $\iota$-th iteration as follows.
\subsection{Alternating optimization for baseband beamforming}
Define
\begin{align}
\ri_{1,\nu}(\bW)&\triangleq r_{\nu}(\bW,\Xii)\nonumber\\
&=\ln\left|\mI_D+[H_{\nu}\Xii\bW_{\nu}]^2\clG^{-1}_{\nu}(\bW,\Xii)\right|\nonumber\\
&=\ln\left|\mI_D+[\clHi_{1,\nu}\bW_{\nu}]^2(\clGi_{1,\nu}(\bW))^{-1}\right|,
\label{bb2}
\end{align}
where
\begin{equation}\label{bb2a}
\clHi_{1,\nu}\triangleq H_{\nu}\Xii,
\end{equation}
and
\begin{equation}\label{bb2a1}
\clGi_{1,\nu}(\bW)\triangleq \sum_{\nu'\neq \nu} [\clHi_{1,\nu}\bW_{\nu'}]^2+\sigma \mI_D.
\end{equation}
Accordingly, define the  minimum rate function
\begin{equation}\label{minrate2}
\fii_1(\bW)\triangleq \min_{\nu\in\clN_u} \ri_{1,\nu}(\bW).
\end{equation}
To generate a baseband beamformer $\Wio$ such that
\begin{equation}\label{wal}
f(\Wio,\Xii) > f(\Wi,\Xii),
\end{equation}
we consider the following subproblem of baseband optimization:
\begin{subequations}\label{bb1}
\begin{eqnarray}
\max_{\bW}\ \fii_1(\bW)\label{bb1a}\\
\mbox{s.t.}\quad \sum_{\nu\in\clN_u}\la (\Xii)^H\Xii,[\bW_{\nu}]^2\ra\leq P,\label{bb1b}
\end{eqnarray}
\end{subequations}
where (\ref{bb1b}) corresponds to the power constraint (\ref{holo10b})
with $\bX$ held fixed at $\Xii$.

As the objective function in (\ref{bb1}) is not concave, the problem is nonconvex. To enable tractable optimization, we construct a concave minorant of the objective function.

Applying the inequality (\ref{fund5}) in Appendix A for
$(\bar{V},\bar{Y})=(\Vi_{1,\nu},\Yi_{1,\nu})\triangleq (\clHi_{1,\nu}\Wi_{\nu}, \clGi_{1,\nu}(\Wi))$ and $(\bV,\bY)\triangleq (\clHi_{1,\nu}\bW_{\nu}, \clGi_{1,\nu}(\bW))$
yields:
\begin{eqnarray}
\ri_{1,\nu}(\bW)&\geq& \tri_{1,\nu}(\bW)\label{bound1}
\end{eqnarray}
with
\begin{eqnarray}
\tri_{1,\nu}(\bW)&\triangleq& \ai_{1,\nu}+2\Re\{\la (\Vi_{1,\nu})^H(\Yi_{1,\nu} )^{-1} \clHi_{1,\nu}\bW_{\nu}\ra\}\nonumber\\
&&-\la \clCi_{1,\nu},
\sum_{\nu'\in\clN_u}[\clHi_{1,\nu}\bW_{\nu'}]^2\ra\label{bb3}\\
&=&\ai_{1,\nu}+2\Re\{\la \clBi_{1,\nu}\bW_{\nu}\ra\}\nonumber\\
&&-\la \tclCi_{1,\nu},
\sum_{\nu'\in\clN_u}[\bW_{\nu'}]^2\ra,\label{bb3e}
\end{eqnarray}
where
\begin{equation}\label{bb4}
\begin{array}{c}
\ai_{1,\nu}\triangleq \ri_{1,\nu}(\Wi)-\la [\Vi_{1,\nu}]^2(\Yi_{1,\nu})^{-1}\ra-\sigma\la\clCi_{1,\nu}\ra,\\
\clCi_{1,\nu}\triangleq (\Yi_{1,\nu})^{-1}-\left(\Yi_{1,\nu}+[\Vi_{1,\nu}]^2\right)^{-1}\succeq 0, \\
\tclCi_{1,\nu}\triangleq (H_{\nu}\Xii)^H\clCi_{1,\nu}H_{\nu}\Xii\succeq 0,\\
\clBi_{1,\nu}\triangleq (\Vi_{1,\nu})^H(\Yi_{1,\nu} )^{-1} \clHi_{1,\nu}.
\end{array}
\end{equation}
Moreover, it is immediate to verify that
\begin{equation}\label{bound2}
\ri_{1,\nu}(\Wi)=\tri_{1,\nu}(\Wi).
\end{equation}
Each $\tri_{1,\nu}(\bW)$ is a concave quadratic function because $\tclCi_{1,\nu}\succeq 0$.
Therefore, the function
\begin{equation}\label{minf1}
\tfii_1(\bW)\triangleq \min_{\nu\in\clN} \tri_{1,\nu}(\bW),
\end{equation} is concave as the pointwise minimum of $\tri_{1,\nu}(\bW)$ \cite{Tuybook}.

From (\ref{bound1}) we have
\begin{eqnarray}
\tfii_{1}(\bW)&\leq&\min_{\nu\in\clN} \ri_{1,\nu}(\bW)\nonumber\\
&=&\fii_1(\bW),\label{bound3}
\end{eqnarray}
and from (\ref{bound2}) it follows that:
\begin{eqnarray}
\tfii_{1}(\Wi)&=&\min_{\nu\in\clN} \ri_{1,\nu}(\Wi)\nonumber\\
&=&\fii_1(\Wi),\label{bound4}
\end{eqnarray}
i.e. $\tfii_{1}(\bW)$ is a tight minorant of $\fii_1(\bW)$ at $\Wi$.

We now generate the subsequent feasible  baseband beamformer $\Wio$  as the optimal solution of the following convex quadratic problem
\begin{equation}\label{bb5}
\max_{\bW}\ \tfii_1(\bW)\quad\mbox{s.t.}\quad (\ref{bb1b}).
\end{equation}
As there are $N_uKD$ variables in this problem, its computational complexity  is on the order of ${\cal O}(N_u^3K^3D^3)$ \cite{peaucelle2002user}.

Finally,
\begin{eqnarray}
\fii_1(\Wio)&\geq&\tfii_1(\Wio)\label{bound5}\\
&>&\tfii_1(\Wi)\label{bound6}\\
&=&\fii_1(\Wi),\label{bound7}
\end{eqnarray}
whenever $\fii_1(\Wio)\neq \fii_1(\Wi)$, which ensures (\ref{wal}). Here, (\ref{bound5}) follows from (\ref{bound3}), (\ref{bound6}) holds because $\Wio$ is the optimal solution of the maximization problem (\ref{bb5}) while $\Wi$ is only feasible, and (\ref{bound7}) follows from (\ref{bound4}).
\subsection{Alternating optimization for holographic beamforming}
Define
\begin{align}\label{bb2}
\ri_{2,\nu}(\bX)\triangleq& r_{\nu}(\Wio,\bX)\nonumber\\
=&\ln\left|\mI_D+[H_{\nu}\bX\Wio_{\nu}]^2(\clGi_{2,\nu}(\bX))^{-1}\right|,
\end{align}
where
\begin{equation}\label{bb2a}
\clGi_{2,\nu}(\bX)\triangleq \sum_{\nu'\neq \nu} [H_{\nu}\bX\Wio_{\nu'}]^2+\sigma \mI_D.
\end{equation}
Accordingly, define the minimum rate function
\begin{equation}\label{bound10}
\fii_2(\bX)\triangleq \min_{\nu\in\clN_u} \ri_{2,\nu}(\bX).
\end{equation}
To generate a holographic beamformer $\Xio$ satisfying
\begin{equation}\label{xal}
f(\Wio,\Xio) > f(\Wio,\Xii),
\end{equation}
we consider the following subproblem of holographic optimization:
\begin{subequations}\label{hb1}
\begin{eqnarray}
\max_{\bX\in\mathbb{R}^{M^2\times K}}\ \fii_2(\bX)\quad\mbox{s.t.}
\quad (\ref{holo10c}), \label{hb1a}\\
||\bX\sqrt{\clAio}||^2\leq P,\label{hb1b}
\end{eqnarray}
\end{subequations}
where
\begin{equation}\label{bb2b}
\clAio\triangleq \sum_{\nu\in\clN_u}[\Wio_{\nu}]^2,
\end{equation}
and (\ref{hb1b}) corresponds to the power constraint (\ref{holo10b})
with $\bW$ held fixed at $\Wio$.

Applying the inequality (\ref{fund5}) in Appendix A for
$(\bar{V},\bar{Y})=(\Vi_{2,\nu},\Yi_{2,\nu})\triangleq(H_{\nu}\Xii\Wio_{\nu}, \clGi_{2,\nu}(\Xii))$ and
$(\bV,\bY)\triangleq (H_{\nu}\bX\Wio_{\nu}, \clGi_{2,\nu}(\bX))$
yields
\begin{eqnarray}
\ri_{2,\nu}(\bX)&\geq&\tri_{2,\nu}(\bX)\label{bound11}
\end{eqnarray}
with
\begin{eqnarray}
\tri_{2,\nu}(\bX)&\triangleq&\ai_{2,\nu}+2\Re\{\la (\Vi_{2,\nu})^H(\Yi_{2,\nu})^{-1} H_{\nu}\bX\Wio_{\nu}\ra\}\nonumber\\
&&-\la \clCi_{2,\nu},
\sum_{\nu'\in\clN_u}[H_{\nu}\bX\Wio_{\nu'}]^2\ra\nonumber\\
&=&\ai_{2,\nu}+2\Re\{\la \clBi_{2\nu}\bX\ra\}\nonumber\\
&&-\la \tclCi_{2,\nu},
[\bX\sqrt{\clAio}]^2\ra, \label{hb3}
\end{eqnarray}
where
\begin{equation}\label{hb4}
\begin{array}{c}
\ai_{2,\nu}\triangleq \ri_{2,\nu}(\Xii)-\la [\Vi_{2,\nu}]^2(\Yi_{2,\nu})^{-1}\ra-\sigma\la\clCi_{2,\nu}\ra,\\
\clCi_{2,\nu}\triangleq (\Yi_{2,\nu})^{-1}-\left(\Yi_{2,\nu}+[\Vi_{2,\nu}]^2\right)^{-1}\succeq  0,\\
\tclCi_{2,\nu}\triangleq (H_{\nu})^H\clCi_{2,\nu}H_{\nu}\succeq 0,\\
\clBi_{2,\nu}\triangleq \Wio_{\nu}(\Vi_{2,\nu})^H(\Yi_{2,\nu})^{-1} H_{\nu}.
\end{array}
\end{equation}
Moreover, it is immediate to verify that
\begin{equation}\label{bound12}
\ri_{2,\nu}(\Xii)=\tri_{2,\nu}(\Xii).
\end{equation}
Each $\tri_{2,\nu}(\bX)$ is a concave quadratic function because $\tclCi_{2,\nu}\succeq 0$.
Therefore, the function
\begin{equation}\label{minf2}
\tfii_2(\bX)\triangleq \min_{\nu\in\clN} \tri_{2,\nu}(\bX),
\end{equation}
is concave as the pointwise minimum of $\tri_{2,\nu}(\bX)$ \cite{Tuybook}.

Similar to (\ref{bound3}) and (\ref{bound4}), it follows from (\ref{bound11}) and (\ref{bound12}) that $\tfii_2(\bX)$ is a tight minorant of $\fii_2(\bX)$ at $\Xii$, i.e.
\begin{equation}\label{bound13}
\tfii_2(\bX)\leq \fii_2(\bX)\quad\forall\ \bX,
\end{equation}
and
\begin{equation}\label{bound14}
\tfii_2(\Xii)=\fii_2(\Xii).
\end{equation}
We then generate the subsequent feasible  holographic beamformer $\Xio$ as the optimal solution of the following convex quadratic problem
\begin{equation}\label{hb5}
\max_{\bX\in\mathbb{R}^{M^2\times K}}\ \tfii_2(\bX)\quad\mbox{s.t.}\quad (\ref{holo10c}), (\ref{hb1b}).
\end{equation}
As there are $M^2K$ variables and $M^2K$ amplitude constraints,
the computational complexity of (\ref{hb5})  is on the order of ${\cal O}((M^2K)^4)={\cal O}(M^8K^4)$ \cite{peaucelle2002user}.\footnote{ $M^2K$ amplitude constraints in (\ref{holo10c}) can be
expressed as $M^2K$ convex quadratic constraints
$\bx_{k,m,m'}(\mu_{k,m,m'}-\bx_{k,m,m'})\geq 0$, $(k,m,m')\in\clK\times \clM\times \clM$ \cite{Tuybook}.
While these constraints do not introduce additional computational difficulties, they certainly contribute to having an increased computational complexity. As a result, the computational complexity of (\ref{hb5}) is not cubic but quartic in  $M^2K$. }

Similarly to (\ref{bound5})-(\ref{bound7}), we have (\ref{xal}) as a consequence of (\ref{bound13}) and (\ref{bound14}) and $\tfii_2(\Xio)>\tfii_2(\Xi)$ whenever $\tfii_2(\Xio)\neq \tfii_2(\Xii)$, since $\Xio$ and $\Xii$ are the optimal solution and a feasible point of the maximization problem (\ref{hb5}), respectively.

\begin{algorithm}[t]
\caption{Convex quadratic solver based algorithm for max-min rate optimization} \label{alg1}
\begin{algorithmic}[1]
\State \textbf{Initialization:} Initialize $(W^{(0)},X^{(0)})$  feasible for (\ref{holo10}). Set $\iota=0$.
\State \textbf{Repeat until convergence:} Generate a baseband beamformer $\Wio$ by
solving the convex quadratic problem (\ref{bb5}) and a holographic beamformer $\Xio$  by solving the convex quadratic problem (\ref{hb5}). Reset $\iota \leftarrow \iota+1$.
\State \textbf{Output} $(W^{opt},X^{opt})=(\Wi,\Xii)$.
\end{algorithmic}
\end{algorithm}
\subsection{The algorithm and its convergence}
Algorithm \ref{alg1} provides the pseudo code of solving the max-min rate problem of (\ref{holo10}). It follows from
(\ref{wal}) and  (\ref{xal})  that we have:
\begin{equation}\label{wzpt1}
f(\Wio,\Xio)> f(\Wi,\Xii)
\end{equation}
as far as $f(\Wio,\Xio)\neq f(\Wi,\Xii)$,
so the sequence $\{(\Wi,\Xii) \}$ of improved feasible points for (\ref{holo10}) converges
to $(\bar{W},\bar{X})$.

\section{Sum-rate driven optmization of holographic beamforming}
By defining the sum-rate function as
\begin{equation}\label{bound15}
f_S(\bW,\bX)\triangleq \sum_{\nu\in\clN_u} r_{\nu}(\bW,\bX),
\end{equation}
it is straightforward to see that the alternating optimization procedure from the previous section can be adapted to solve the following sum-rate problem:
\begin{equation}\label{sumrate}
\max_{\bW,\bX}\ f_S(\bW,\bX)\quad\mbox{s.t.}\quad
(\ref{holo10b}), (\ref{holo10c}).
\end{equation}
Alternating baseband optimization  relies on the optimal solution of the convex quadratic
problem (\ref{bb5}) with $\tfii_1(\bW)$ defined by $\sum_{\nu\in\clN} \tri_{1,\nu}(\bW)$ instead of (\ref{minf1}). As seen below, this problem admits expressing the optimal solution in closed-form.
Furthermore, alternating holographic optimization  is based on the optimal solution of the convex quadratic problem (\ref{hb5}) with $\tfii_2(\bX)$ defined by $\sum_{\nu\in\clN} \tri_{2,\nu}(\bX)$ instead of (\ref{minf2}). Its computational complexity remains  high at  ${\cal O}(M^8K^4)$ due to having multiple amplitude constraints.

To handle the multitude of amplitude constraints (\ref{holo10c}) in (\ref{sumrate}), we utilize the widely used penalized optimization framework  \cite{Betal06,PTKD12} to incorporate them into the optimization objective as
\begin{equation}\label{bound16}
f_{S,\rho}(\bW,\bX,\bchi)\triangleq f_{S}(\bW,\bX)-\rho||{\sf vec}(\bX)-\bchi||^2,
\end{equation}
where $\rho>0$ is the penalty parameter that drives the holographic beamformer $\bX$ toward feasibility with respect to constraint (\ref{holo10c}) upon convergence.
This yields the following penalized optimization problem:
\begin{subequations}\label{gm1}
\begin{eqnarray}
\max_{\bW, \bX, \bchi} f_{S,\rho}(\bW,\bX,\bchi)\\
\mbox{s.t.}\quad (\ref{holo10b}), \label{gm1a}\\
\bchi\triangleq {\sf vec}\left(\begin{bmatrix}\bchi_1&\dots&\bchi_K\end{bmatrix}\right)\in\mathbb{R}^{M^2K},\nonumber\\
\bchi_k\triangleq {\sf vec}\left([\bchi_{k,m,m'}]_{(m,m')\in\clM\times\clM}\right)\in\mathbb{R}^{M^2},\label{gm1b}\\
\bchi_{k,m,m'}\in [0,\mu_{k,m,m'}], (k,m,m')\in\clK\times\clM\times\clM,\label{gm1c}
\end{eqnarray}
\end{subequations} The penalized optimization problem (\ref{gm1}) eliminates the explicit amplitude constraints (\ref{holo10c}) imposed on $\bX$. As the penalty term approaches zero, a solution feasible for (\ref{gm1}) is also feasible for (\ref{sumrate}). More importantly, unlike (\ref{sumrate}), the penalized reformulation (\ref{gm1}) enables computationally tractable alternating optimization in $\bX$ and $\bchi$ with scalable complexity. The interested reader is referred to \cite[Chapter 16]{Betal06} for further insights into penalty optimization.

We now propose an alternating optimization algorithm that iterates by evaluating closed-form expressions for
addressing (\ref{gm1}).

Let $(W^{(0)}, X^{(0)}, \chi^{(0)})$ be  an initial feasible point
and  $(\Wi, \Xii,\chii)$ be the feasible point for (\ref{gm1})
found from the $(\iota-1)$-st iteration. Alternating optimization in each of
$\bW$, $\bX$ and $\bchi$ is progressed as follows.
\subsection{Alternating optimization for baseband beamforming}
Alternating baseband optimization  aims for generating $\Wio$ to ensure that
\begin{eqnarray}\label{wal1}
&&f_{S,\rho}(\Wio,\Xii,\chii)> f_{S,\rho}(\Wi,\Xii,\chii)\nonumber\\
&\Leftrightarrow&
f_{S}(\Wio,\Xii)>f_{S}(\Wi,\Xii).
\end{eqnarray}
To achieve this, we define the sum-rate function
\begin{equation}\label{bound18}
\fii_{1,S}(\bW)\triangleq \sum_{\nu\in\clN_u}\ri_{1,\nu}(\bW),
\end{equation}
where  $\ri_{1,\nu}(\bW)$ is
 defined in (\ref{bb2})-(\ref{bb2a1}), and consider the following sum-rate maximization problem, instead of the max-min rate problem (\ref{bb1}):
\begin{equation}\label{gbb1}
\max_{\bW}\ \fii_{1,S}(\bW)\quad
\mbox{s.t.}\quad (\ref{bb1b}).
\end{equation}
With the tight minorant $\tri_{1,\nu}(\bW)$ of $\ri_{1,\nu}(\bW)$
as defined in
(\ref{bb3e})-(\ref{bb4}), we obtain
\begin{eqnarray}
\fii_{1,S}(\bW)&\geq&\tfi_{1,S}(\bW)\label{bound19}
\end{eqnarray}
with
\begin{eqnarray}
\tfi_{1,S}(\bW)&\triangleq&\sum_{\nu\in\clN_u}\tri_{1,\nu}(\bW)\nonumber\\
&=&\sum_{\nu\in\clN_u}\left[\ai_{1,\nu}+2\Re\{\la \clBi_{1,\nu}\bW_{\nu}\ra\}\right.\nonumber\\
&&\left.-\la \tclCi_{1,\nu},\sum_{\nu'\in\clN_u}[\bW_{\nu'}]^2\ra
\right]\nonumber\\
&=&\sum_{\nu\in\clN_u}\ai_{1,\nu}+\sum_{\nu\in\clN_u}\left[2\Re\{\la \clBi_{1,\nu}\bW_{\nu}\ra\}\right.\\
&&\left.-\la \tclCi_1, [\bW_{\nu}]^2\ra\right],\label{gbb2}
\end{eqnarray}
where:
\begin{equation}\label{gbb3}
\tclCi_1\triangleq \sum_{\nu\in\clN_u}\tclCi_{1,\nu}.
\end{equation}
In addition,
\begin{equation}\label{bound20}
\tfi_{1,S}(\Wi)= \fii_{1,S}(\Wi).
\end{equation}
The subsequent baseband beamformer $\Wio$ ensuring (\ref{wal1}) is then obtained as the optimal solution of
\begin{equation}\label{gbb4}
\max_{\bW} \tfi_{1,S}(\bW)\quad
\mbox{s.t.}\quad (\ref{bb1b}),
\end{equation}
which admits the following closed-form solution:
\begin{equation}\label{gbb5}
\Wio_{\nu}=\begin{cases}\begin{array}{l}\!\!\!\!(\tclCi_1)^{-1}(\clBi_{1,\nu})^H\\
\!\!\!\!\mbox{if}\ \ds\sum_{\nu\in\clN_u}\la (\Xii)^H\Xii,[(\tclCi_1)^{-1}(\clBi_{1,\nu})^H]^2\ra\!\leq\! P,\cr
\!\!\!\!(\tclCi_1+\tau(\Xii)^H\Xii)^{-1}(\clBi_{1,\nu})^H\\
\!\!\!\!\mbox{otherwise},
\end{array}
\end{cases}
\end{equation}
where $\tau>0$ is found by bisection such that\footnote{{\bf Bisection procedure.} Define $g(\tau)\!\!\triangleq\!\!\sum_{\nu\in\clN_u}\!\!\la (\Xii)^H\Xii,[(\tclCi_1+\tau(\Xii)^H\Xii)^{-1}(\clBi_{1,\nu})^H]^2\ra$. Find bounds $\tau_l$ and $\tau_u$ satisfying $g(\tau_l)>P>g(\tau_u)$. Then, set $\tau=(\tau_l+\tau_u)/2$ and compute $g(\tau)$. Terminate the procedure whenever $g(\tau)\approx P$. Otherwise, update $\tau_l\leftarrow \tau$ if $g(\tau)>P$ or update $\tau_u\leftarrow \tau$ if $g(\tau)<P$.}
\begin{equation}
\ds\sum_{\nu\in\clN_u}\!\!\la (\Xii)^H\Xii,[(\tclCi_1+\tau(\Xii)^H\Xii)^{-1}(\clBi_{1,\nu})^H]^2\ra\!=\! P.
\end{equation}
The computational complexity of expression (\ref{gbb5}) is on the order of ${\cal O}(N_uKD)$.

Similarly to (\ref{bound5})-(\ref{bound7}), we have (\ref{wal1}) as a consequence of (\ref{bound19}) and (\ref{bound20}) and $\tfi_{1,S}(\Wio)>\tfi_{1,S}(\Wi)$ whenever $\tfi_{1,S}(\Wio)\neq \tfi_{1,S}(\Wi)$, since $\Wio$ and $\Wi$ are the optimal solution and a feasible point of the maximization problem (\ref{gbb4}), respectively.
\subsection{Alternating optimization for holographic beamforming}
Holographic alternating optimization  aims for generating $\Xio$ for ensuring that
\begin{equation}
f_{S,\rho}(\Wio,\Xio,\chii)>f_{S,\rho}(\Wio,\Xii,\chii).\label{xal1}
\end{equation}
To this end, we define the sum-rate function as
\begin{equation}\label{bound21}
\fii_{2,S}(\bX)\triangleq \sum_{\nu\in\clN_u} \ri_{2,\nu}(\bX)
\end{equation}
where $\ri_{2,\nu}(\bX)$ is given in (\ref{bb2}). The penalized sum-rate is then
\begin{eqnarray}
\fii_{2,S,\rho}(\bX)\triangleq \fii_{2,S}(\bX)-\rho||{\sf vec}(\bX)-\chii||^2,
\end{eqnarray}
leading to the following optimization problem:
\begin{align}\label{ghb1}
\max_{\bX}\ \fii_{2,S,\rho}(\bX)\quad \mbox{s.t.}\quad (\ref{hb1b}).
\end{align}
With the tight minorant $\tri_{2,\nu}(\bX)$ of $r_{2,\nu}(\bX)$ as
defined in (\ref{hb3})-(\ref{hb4}), we obtain
\begin{eqnarray}
\fii_{2,S}(\bX)&\geq&\tfi_{2,S}(\bX)\label{bound22}
\end{eqnarray}
with
\begin{eqnarray}
\tfi_{2,S}(\bX)&\triangleq&\sum_{\nu\in\clN_u}\tri_{2,\nu}(\bX)\nonumber\\
&=&\sum_{\nu\in\clN_u}\ai_{2,\nu}+2(\bi_2)^T{\sf vec}(\bX)\nonumber\\
&&-{\sf vec}^T(\bX) \tclDi_2{\sf vec}(\bX)\label{vec5}
\end{eqnarray}
where
\begin{equation}\label{vec2}
\bi_2\triangleq {\sf vec}\left(\Re\{(\tclBi_2)^T\}\right)\in\mathbb{R}^{M^2K}\ni,
\end{equation}
and
\begin{equation}\label{vec4}
\tclDi_2\triangleq \Re\{ (\clAio)^T\otimes \tclCi_2 \}
\end{equation}
for  $\clAio$ and $(\ai_{2,\nu},\clBi_{2,\nu},\tclCi_{2,\nu})$  defined in (\ref{bb2b})
and (\ref{hb4}), and
\begin{equation}\label{ghb3}
\begin{array}{c}
\tclBi_2\triangleq \sum_{\nu\in\clN_u}\clBi_{2,\nu},\\
\tclCi_2\triangleq \sum_{\nu\in\clN_u}\tclCi_{2,\nu}.
\end{array}
\end{equation}
The derivation for (\ref{vec5}) is provided in Appendix B.

In addition,
\begin{equation}\label{bound23}
\tfi_{2,S}(\Xii)= \fii_{2,S}(\Xii).
\end{equation}
By further defining
\begin{eqnarray}
\tfi_{2,S,\rho}(\bX)&\triangleq&\tfi_{2,S}(\bX)-\rho||{\sf vec}(\bX)-\chii||^2\nonumber\\
&=&\sum_{\nu\in\clN_u}\ai_{2,\nu}+2(\bi_2)^T{\sf vec}(\bX)\nonumber\\
&&-{\sf vec}^T(\bX) \tclDi_2{\sf vec}(\bX)\nonumber\\
&& -\rho||{\sf vec}(\bX)-\chii||^2,\label{bound24}
\end{eqnarray}
it follows from (\ref{bound22}) and (\ref{bound24}) that
\begin{equation}\label{bound25}
\tfi_{2,S,\rho}(\bX)\leq \fii_{2,S,\rho}(\bX)\ \forall\ \bX,
\end{equation}
and
\begin{equation}\label{bound26}
\tfi_{2,S,\rho}(\Xii)= \fii_{2,S,\rho}(\Xii).
\end{equation}
In other words, $\tfi_{2,S,\rho}(\bX)$ is a tight minorant of the objective function $\fii_{2,S,\rho}(\bX)$ in (\ref{ghb1}).

The power constraint (\ref{hb1b}) in (\ref{ghb1}) can be written as
\begin{equation}\label{vec6}
{\sf vec}^T(\bX) \tclDi_1{\sf vec}(\bX)\leq P,
\end{equation}
with
\begin{equation}\label{vec7}
\tclDi_1\triangleq  \Re\{(\clAio)^T\}\otimes I_{M^2}.
\end{equation}
The subsequent holographic beamformer $\Xio$ is then obtained as the optimal solution of
\begin{align}\label{ghb4}
\max_{{\sf vec}(\bX)}\ \tfi_{2,S,\rho}(\bX)  \quad\mbox{s.t.}
\quad (\ref{vec6}),
\end{align}
which admits the closed-form solution of
\begin{align}\label{1ghb5}
&{\sf vec}(\Xio)\nonumber\\
&=\begin{cases}\begin{array}{l}\!\!\!\!(\tclDi_2+\rho I_{KM^2})^{-1} (\bi_2+\rho\chii)\\\
\!\!\!\!\mbox{if}\ ||\sqrt{\tclDi_1}(\tclDi_2+\rho I_{KM^2})^{-1}(\bi_2+\rho\chii)||^2\leq P,\cr
\!\!\!\!(\tclDi_2+\rho I_{KM^2}+\tau \tclDi_1)^{-1}(\bi_2+\rho\chii)\\
\!\!\!\!\mbox{otherwise},
\end{array}
\end{cases}
\end{align}
where $\tau>0$ is found by bisection so that
\begin{equation}\label{1ghb6}
||\sqrt{\tclDi_1}(\tclDi_2+\rho I_{KM^2}+\tau \tclDi_1)^{-1}  (\bi_2
+\rho\chii)||^2=P.
\end{equation}
The computational complexity of expression  (\ref{1ghb5}) is on the order of ${\cal O}(M^2K)$.

Similarly to (\ref{bound5})-(\ref{bound7}), we have (\ref{xal1}) as a consequence of (\ref{bound25}) and (\ref{bound26}) and $\tfi_{2,S,\rho}(\Xio)>\tfi_{2,S,\rho}(\Xii)$ whenever $\tfi_{2,S,\rho}(\Xio)\neq \tfi_{2,S,\rho}(\Xii)$, since $\Xio$ and $\Xii$ are the optimal solution and a feasible point of the maximization problem (\ref{ghb4}), respectively.
\subsection{Amplitude optimization and control}
The amplitude optimization and control aims for generating $\chio$ so that
\begin{align}
&f_{S,\rho}(\Wio,\Xio,\chio)>\nonumber\\
&f_{S,\rho}(\Wio,\Xio,\chii)\label{cal1}\\
\Leftrightarrow& ||\xio-\chiio||^2< ||\xio-\chii||^2,\label{cal1a}
\end{align}
for $\xio\triangleq {\sf vec}(\Xio)$. Hence, $\chio$ is obtained as the optimal solution of
\begin{equation}\label{cal1b}
\min_{\bchi} ||\xio-\bchi||^2\quad\mbox{s.t.}\quad (\ref{gm1c}),
\end{equation}
i.e.\footnote{The notation $\arg\min$ returns the argument (value of the variable) that minimizes the objective function of the corresponding optimization problem.}
\begin{align}
\chio_{k,m,m'}=&\arg\min_{\bchi_{k,m,m'}\in [0,\mu_{k,m,m'}]}(\xio_{k,m,m'}-\bchi_{k,m,m'})^2\nonumber\\
=&\begin{cases}\begin{array}{ll}\xio_{k,m,m'}&\mbox{if}\quad 0\leq \xio_{k,m,m'}\leq \mu_{k,m,m'}\cr
0&\mbox{if}\quad \xio_{k,m,m'}<0\cr
\mu_{k,m,m'}&\mbox{if}\quad \xio_{k,m,m'}>\mu_{k,m,m'},
\end{array}
\end{cases}\nonumber\\
&\quad (k,m,m')\in\clK\times\clM\times\clM.\label{mgm1}
\end{align}
The computational complexity of expression (\ref{mgm1}) is on the order of ${\cal O}(M^2K)$.
\subsection{The algorithm and its convergence}
Algorithm \ref{alg2} provides the pseudo code for computing the problem (\ref{gm1}).
It follows from (\ref{wal1}), (\ref{xal1}) and (\ref{cal1}) that
\begin{equation}\label{wxc1}
f_{S,\rho}(\Wio,\Xio,\chio)> f_{S,\rho}(\Wi,\Xii,\chii),
\end{equation}
as far as $f_{S,\rho}(\Wio,\Xio,\chio)\neq f_{S,\rho}(\Wi,\Xii,\chii)$, so the sequence
$\{(\Wi,\Xii,\chii)\}$ converges to $(\bar{W},\bar{X},\bar{\chi})$. Furthermore, by choosing a sufficiently large $\rho>0$, we can ensure that  $||{\sf vec}(\Xii)-\chii||^2\rightarrow 0$ as $\iota\rightarrow\infty$, which means that
$(\bar{W},\bar{X})$ is a feasible point for   (\ref{sumrate}), which turns out to be at least a local solution \cite{Betal06}.
\begin{algorithm}[!t]
\caption{Scalable-complexity algorithm for sum-rate optimization} \label{alg2}
\begin{algorithmic}[1]
\State \textbf{Initialization:} Initialize $(X^{(0)},W^{(0)},\chi^{(0)})$  feasible for (\ref{gm1}). Set $\iota=0$.
\State \textbf{Repeat until convergence:} Generate  $\Wio$ by the closed-form
(\ref{gbb5}), $\Xio$  by the closed form (\ref{1ghb5}), and $\chio$ by the closed form
(\ref{mgm1}). Reset $\iota \leftarrow \iota+1$.
\State \textbf{Output} $(X^{opt}, W^{opt})=(\Xii,\Wi)$.
\end{algorithmic}
\end{algorithm}
\section{Surrogate  optimization having scalable-complexity for enhancing both the min-rate and sum-rate}
It is plausible that one can utilize the following penalized optimization formulation to address the max-min rate optimization problem (\ref{holo10}):
\begin{align}\label{pm1}
\max_{\bW, \bX, \bchi}\left[f(\bW,\bX)-\rho||{\sf vec}(\bX)-\bchi||^2\right]\nonumber\\
\mbox{s.t.}\quad (\ref{holo10b}), (\ref{gm1b}), (\ref{gm1c}).
\end{align}
However, due to the non-smooth nature of the function $f(\bW,\bX)$, this penalized optimization does not offer benefits, since its computational complexity remains as high as that of Algorithm \ref{alg1}.

Following \cite{Tuaetal24,Zhuetal23twc,Zhuetal24twc,Cheetal24tcom} we scale each rate function
$r_{\nu}(\bW,\bX)$ as
\begin{equation}\label{rnc}
r_{\nu,c}(\bW,\bX)\triangleq \ln\left|\mI_D+\frac{1}{c}[H_{\nu}\bX\bW_{\nu}]^2\clG^{-1}_{\nu}(\bW,\bX)\right|,
\end{equation}
with $0<c\leq 1$. Here, $r_{\nu,c}(\bW,\bX)>r_{\nu,1}(\bW,\bX)=r_{\nu}(\bW,\bX)$ and
$r_{\nu,c}(\bW,\bX)\rightarrow +\infty$ as $c\rightarrow 0$. Maximizing the non-smooth function
$f_c(\bW,\bX)\triangleq \min_{\nu\in\clN_u}r_{\nu,c}(\bW,\bX)$ enhances $f(\bW,\bX)$. More importantly, $f_c(\bW,\bX)$ satisfies the following two-sided inequality
\begin{equation}\label{rnc1}
f_c(\bW,\bX) >  -\ln\left|\sum_{\nu\in\clN_u}\Pi_{\nu}(\bW,\bX) \right| >   f_c(\bW,\bX)-\ln N_u,
\end{equation}
where $\Pi_{\nu}(\bW,\bX)$ is defined in (\ref{rnc3}) (see the equation at the bottom of the next page).
\begin{figure*}[b]
\vspace{-0.4cm}
\hrulefill
\begin{align}
\Pi_{\nu}(\bW,\bX)\triangleq&\left(\mI_D+\frac{1}{c}(H_{\nu}\bX\bW_{\nu})^H\clG^{-1}_{\nu}(\bW,\bX)(H_{\nu}\bX\bW_{\nu})
\right)^{-1}\label{rnc2}\\
=&\mI_D-(H_{\nu}\bX\bW_{\nu})^H\left([H_{\nu}\bX\bW_{\nu}]^2+
c\clG_{\nu}(\bW,\bX)\right)^{-1}(H_{\nu}\bX\bW_{\nu}).\label{rnc3}
\end{align}
\end{figure*}
As such, the smooth  function $-\ln\left|\sum_{\nu\in\clN_u}\Pi_{\nu}(\bW,\bX) \right|$ may be viewed as a soft approximation of the non-smooth pointwise minimum function $f_c(\bW,\bX)$
\cite{EMMZ20}. Instead of maximizing the non-smooth objective function $f_c(\bW,\bX)$, our goal is now to maximize its soft smooth approximation $-\ln\left|\sum_{\nu\in\clN_u}\Pi_{\nu}(\bW,\bX) \right|$ or equivalently, to minimize $\ln\left|\sum_{\nu\in\clN_u}\Pi_{\nu}(\bW,\bX) \right|$ subject to the constraints (\ref{holo10b}) and (\ref{holo10c}):
\begin{equation}\label{osm}
\min_{\bW,\bX} \ln |\Pi(\bW,\bX)|\quad\mbox{s.t.}\quad
(\ref{holo10b}), (\ref{holo10c}),
\end{equation}
where $\Pi(\bW,\bX)$ is defined in (\ref{smm2}) of the next page.

To solve (\ref{osm}) we introduce the penalized objective
\begin{equation}\label{defphi}
\varphi(\bW,\bX,\bchi)\triangleq \ln |\Pi(\bW,\bX)|+\rho||{\sf vec}(\bX)-\bchi||^2,
\end{equation}
and consider the smooth optimization problem
\begin{equation}\label{smm1}
\min_{\bW,\bX,\bchi}\ \varphi(\bW,\bX,\bchi)\quad\mbox{s.t.}\quad (\ref{holo10b}), (\ref{gm1b}), (\ref{gm1c}).
\end{equation}
We will develop an algorithm based on closed-form expressions to solve (\ref{smm1}); specifically, closed-form updates for $\bW$, $\bX$ and $\bchi$ are derived within an alternating-optimization framework, yielding an efficient implementation.
\begin{figure*}[b]
\vspace{-0.4cm}
\hrulefill
\begin{align}\label{smm2}
\Pi(\bW,\bX)\triangleq& \sum_{\nu\in\clN_u}\Pi_{\nu}(\bW,\bX)\nonumber\\
=& \sum_{\nu\in\clN_u}\left[ \mI_D-(H_{\nu}\bX\bW_{\nu})^H\left([H_{\nu}\bX\bW_{\nu}]^2+c\clG_{\nu}(\bW,\bX) \right)^{-1}(H_{\nu}\bX\bW_{\nu})\right].
\end{align}
\end{figure*}

Let $(W^{(0)}, X^{(0)}, \chi^{(0)})$ be  an initial feasible point
and  $(\Wi, \Xii,\chii)$ be the feasible point for (\ref{smm1})
found from the $(\kappa-1)$-st iteration. The alternating optimization for $\bW$, $\bX$, and $\bchi$ then proceeds as follows.
\subsection{Alternating optimization for baseband beamforming}
Baseband alternating optimization  aims for generating $\Wio$ so that
\begin{eqnarray}\label{swal1}
&&\varphi(\Wio,\Xii,\chii)< \varphi(\Wi,\Xii,\chii)\nonumber\\
&\Leftrightarrow&\ln |\Pi(\Wio,\Xii)|<\ln |\Pi(\Wi,\Xii)|.
\end{eqnarray}
To achieve this, we define $\Pii_1(\bW)$ as in (\ref{bsm5}) on the next-to-next page, and
consider the following optimization problem:
\begin{equation}\label{bsm3}
	\min_{\bW}\ \ln|\Pii_1(\bW)|\quad \mbox{s.t.}\quad (\ref{bb1b}).
\end{equation}
\begin{figure*}[b]
\vspace{-0.4cm}
\hrulefill
\begin{align}\label{bsm5}
\Pii_1(\bW)\triangleq& \Pi(\bW,\Xii)\nonumber\\
=& \sum_{\nu\in\clN_u}\left[ \mI_D-(\clHi_{1,\nu}\bW_{\nu})^H
\left[[\clHi_{1,\nu}\bW_{\nu}]^2+c\left(\sum_{\nu'\neq \nu} [\clHi_{1,\nu}\bW_{\nu'}]^2+\sigma \mI_D \right) \right]^{-1}
(\clHi_{1,\nu}\bW_{\nu})\right],
\end{align}
where $\clHi_{1,\nu}$ is defined from (\ref{bb2a}).
\end{figure*}
Using the inequality (\ref{ap6}) in Appendix A for
\[
\begin{array}{lll}
(\bV_{\nu},\bY_{\nu})&=&(\clHi_{1,\nu}\bW_{\nu}, [\clHi_{1,\nu}\bW_{\nu}]^2\\
&&+c(\sum_{\nu'\neq \nu} [\clHi_{1,\nu}\bW_{\nu'}]^2+\sigma \mI_D)),\nu\in\clN_u,
\end{array}
\]
and
\[
\begin{array}{lll}
(\bar{V}_{\nu},\bar{Y}_{\nu})&=&(\Vi_{1,\nu},\Yi_{1,\nu})\\
&\triangleq& (\clHi_{1,\nu}\Wi_{\nu},[\clHi_{1,\nu}\Wi_{\nu}]^2 \nonumber\\
&&+c(\sum_{\nu'\neq \nu} [\clHi_{1,\nu}\Wi_{\nu'}]^2+\sigma \mI_D)),\nu\in\clN_u,
\end{array}
\]
yields
\begin{eqnarray}
\ln|\Pii_1(\bW)|&\leq& \fii_{1}(\bW)\label{bound30}
\end{eqnarray}
with
\begin{eqnarray}
 \fii_{1}(\bW)&\triangleq& \ai_1-2\sum_{\nu\in\clN_u}\Re\{\la\clBi_{1,\nu}\bW_{\nu}\ra\}\nonumber\\
&&+\sum_{\nu\in\clN_u}\la\clCi_{1,\nu},[\clHi_{1,\nu}\bW_{\nu}]^2\nonumber\\
&&+c\sum_{\nu'\in\clN_u\setminus\{\nu\}}\!\![\clHi_{1,\nu}\bW_{\nu'}]^2\ra\label{bsm6}\\
&=&\ai_1-2\!\sum_{\nu\in\clN_u}\Re\{\la\clBi_{1,\nu}\bW_{\nu}\ra\}\nonumber\\
&&+\sum_{\nu\in\clN_u}\la \tclCi_{1,\nu},[\bW_{\nu}]^2\ra, \label{bsm8}
\end{eqnarray}
where
\begin{subequations}\label{bsm7}
\begin{align}
\ai_1\triangleq& \ln|\Pii_1(\Wi)|\nonumber\\
&+ \sum_{\nu\in\clN_u}\la (\Pii_1(\Wi))^{-1}(\Vi_{1,\nu})^H(\Yi_{1,\nu})^{-1}\Vi_{1,\nu}\ra\nonumber\\
&+c\sum_{\nu\in\clN_u}\la \clCi_{1,\nu}\ra,	\label{bsm7a}\\
\clBi_{1,\nu}\triangleq& (\Pii_1(\Wi))^{-1}(\Vi_{1,\nu})^H(\Yi_{1,\nu})^{-1}\clHi_{1,\nu},	\label{bsm7b}\\
\clCi_{1,\nu}\triangleq& (\Yi_{1,\nu})^{-1}\Vi_{1,\nu}(\Pii_1(\Wi))^{-1}(\Vi_{1,\nu})^H(\Yi_{1,\nu})^{-1}\succeq 0,\label{bsm7c}
\end{align}
\end{subequations}
and
\begin{align}\label{bsm9}
	\tclCi_{1,\nu}\triangleq& (\clHi_{1,\nu})^H \clCi_{1,\nu}\clHi_{1,\nu}\nonumber\\
	&+c\sum_{\nu'\in\clN_u\setminus\{\nu\}}(\clHi_{1,\nu'})^H \clCi_{1,\nu'}\clHi_{1,\nu'}\succeq 0, \nu\in\clN_u.
\end{align}
In addition,
\begin{equation}\label{bound31}
\ln|\Pii_1(\Wi)|=\fii_{1}(\Wi).
\end{equation}
The function $\fii_{1}(\bW)$ is convex quadratic,  because $\tclCi_{1,\nu}\succeq 0$.
The subsequent baseband beamformer $\Wio$  is  obtained as the optimal solution of the following convex quadratic problem:
\begin{equation}\label{bsm10}
\min_{\bW} \fii_{1}(\bW)\quad
\mbox{s.t.}\quad (\ref{bb1b}),
\end{equation}
which admits the solution in the following closed-form
\begin{equation}\label{bsm11}
\Wio_{\nu}=\begin{cases}\begin{array}{l}\!\!\!\!(\tclCi_{1,\nu})^{-1}(\clBi_{1,\nu})^H\\
\!\!\!\!\mbox{if}\ \ds\sum_{\nu\in\clN_u}\!\!\la (\Xii)^H\Xii,[(\tclCi_{1,\nu})^{-1}(\clBi_{1,\nu})^H]^2\ra\!\leq\! P,\cr
\!\!\!\!(\tclCi_{1,\nu}+\tau(\Xii)^H\Xii)^{-1}(\clBi_{1,\nu})^H\\
\!\!\!\!\mbox{otherwise},
\end{array}
\end{cases}
\end{equation}
where $\tau>0$ is found by bisection so that
\begin{equation}
\ds\sum_{\nu\in\clN_u}\!\!\la (\Xii)^H\Xii,[(\tclCi_{1,\nu}+\tau(\Xii)^H\Xii)^{-1}(\clBi_{1,\nu})^H]^2\ra\!=\! P.
\end{equation}
The computational complexity of expression (\ref{bsm11}) is on the order of ${\cal O}(N_uKD)$.

Similarly to (\ref{bound5})-(\ref{bound7}), we have (\ref{swal1}) as a consequence of (\ref{bound30}) and (\ref{bound31}) and $\fii_{1}(\Wio)<\fii_{1}(\Wi)$ whenever $\fii_{1}(\Wio)\neq \fii_{1}(\Wi)$, since $\Wio$ and $\Wi$ are the optimal solution and a feasible point of the minimization problem (\ref{bsm10}), respectively.
\subsection{Alternating optimization for holographic beamforming}
Holographic alternating optimization  aims for generating $\Xio$ such that
\begin{align}
&\varphi(\Wio,\Xio,\chii)<\varphi(\Wio,\Xii,\chii)\label{sxal1}\\
\Leftrightarrow&
\ln |\Pi(\Wio,\Xio)|+\rho||{\sf vec}(\Xio)-\chii||^2
< \nonumber\\
&\ln |\Pi(\Wio,\Xii)|+\rho||{\sf vec}(\Xii)-\chii||^2.\label{sxal1a}
\end{align}
To this end, we define $\Pii_2(\bX)$ as in (\ref{gsm2}) at the bottom of the next page, and
introduce
\begin{equation}\label{phi2}
\varphii_2(\bX)\triangleq  \ln|\Pii_2(\bX)| +\rho||\bx-\chii||^2
\end{equation}
to consider the following optimization problem
\begin{equation}\label{gsm1}
\min_{\bX\in\mathbb{R}^{M^2\times K}} \varphii_2(\bX)\quad\mbox{s.t.}
\quad (\ref{vec6}).
\end{equation}

\begin{figure*}[b]
\vspace{-0.4cm}
\hrulefill
\begin{align}
\Pii_2(\bX)\triangleq&\Pi(\Wio,\bX)\nonumber\\
=& \sum_{\nu\in\clN_u}\left[ \mI_D-(H_{\nu}\bX\Wio_{\nu})^H
\left[[H_{\nu}\bX\Wio_{\nu}]^2+c\left(\sum_{\nu'\neq \nu} [H_{\nu}\bX\Wio_{\nu'}]^2+\sigma \mI_D \right) \right]^{-1}
(H_{\nu}\bX\Wio_{\nu})\right].\label{gsm2}
\end{align}
\end{figure*}
Applying the inequality (\ref{ap6}) in Appendix A for
\begin{align}
(\bV_{\nu},\bY_{\nu})=&
(H_{\nu}\bX\Wio_{\nu},[H_{\nu}\bX\Wio_{\nu}]^2\nonumber\\
&+c(\sum_{\nu'\neq \nu} [H_{\nu}\bX\Wio_{\nu'}]^2+\sigma \mI_D)),\nu\in\clN_u,\nonumber
\end{align}
and
\begin{align}
(\bar{V}_{\nu},\bar{Y}_{\nu})=&(\Vi_{2,\nu},\Yi_{2,\nu})\nonumber\\
\triangleq &(H_{\nu}\Xii\Wio_{\nu}, [H_{\nu}\Xii\Wio_{\nu}]^2 \nonumber\\
&+c(\sum_{\nu'\neq \nu} [H_{\nu}\Xii\Wio_{\nu'}]^2+\sigma \mI_D)),
\nu\in\clN_u,\nonumber
\end{align}
yields
\begin{eqnarray}
\ln|\Pii_2(\bX)|&\leq &\fii_{2}(\bX)\label{bound40}
\end{eqnarray}
with
\begin{eqnarray}
\fii_{2}(\bX)&\triangleq&\ai_2-2\Re\{\la(\sum_{\nu\in\clN_u}\Wio_{\nu}\clBi_{2,\nu}H_{\nu})\bX\ra\}\nonumber\\
&&+\sum_{\nu\in\clN_u}\la\tclCi_{2,\nu},[\bX\Wio_{\nu}]^2\nonumber\\
&&+c\sum_{\nu'\in\clN_u\setminus\{\nu\}}[\bX\Wio_{\nu'}]^2\ra\label{gsm3}\\
&=&\ai_{2}-2(\bi_2)^T{\sf vec}(\bX)\nonumber\\
&&+{\sf vec}(\bX)^T \tclDi_2{\sf vec}(\bX),\label{gvec5}
\end{eqnarray}
where
\begin{equation}\label{gvec2}
\bi_2\triangleq {\sf vec}\left(\Re\{(\tclBi_2)^T\}\right)\in\mathbb{R}^{M^2K},
\end{equation}
and
\begin{equation}\label{gvec5a}
\tclDi_2\triangleq \sum_{\nu\in\clN_u}\tclDi_{2,\nu}\succeq 0.
\end{equation}
for
\begin{subequations}\label{gsm5}
\begin{align}
\ai_2\triangleq& \ln|\Pii_2(\Xii)| \nonumber\\
&+\sum_{\nu\in\clN_u}\la (\Pii_2(\Xii))^{-1}(\Vi_{2,\nu})^H(\Yi_{2,\nu})^{-1}\Vi_{2,\nu}\ra\nonumber\\
&+c\sum_{\nu\in\clN_u}\la \clCi_{2,\nu}\ra,	\label{gsm5a}\\
\clBi_{2,\nu}\triangleq& (\Pii_2(\Xii))^{-1}(\Vi_{2,\nu})^H(\Yi_{2,\nu})^{-1}, \label{gsm5b}\\
\clCi_{2,\nu}\triangleq& (\Yi_{2,\nu})^{-1}\Vi_{2,\nu}(\Pii_2(\Xii))^{-1}(\Vi_{2,\nu})^H(\Yi_{2,\nu})^{-1}\succeq 0,\label{gsm5c}
\end{align}
\end{subequations}
and
\begin{equation}\label{gsm6}
\begin{array}{c}
\tclBi_2\triangleq \sum_{\nu\in\clN_u}\Wio_{\nu}\clBi_{2,\nu}H_{\nu},\\
	\tclCi_{2,\nu}\triangleq H^H_{\nu} \clCi_{2,\nu}H_{\nu}\succeq 0, \nu\in\clN_u,\\
\clAio_{\nu}\triangleq [\Wio_{\nu}]^2+c\sum_{\nu'\in\clN_u\setminus\{\nu\}}[\Wio_{\nu'}]^2, \nu\in\clN_u,
\end{array}
\end{equation}
and
\begin{equation}\label{gvec4}
\tclDi_{2,\nu}\triangleq \Re\{ (\clAio_{\nu})^T\otimes \tclCi_{2,\nu}\}.
\end{equation}
The derivation of (\ref{gvec5}) is provided in Appendix C.

In addition,
\begin{equation}\label{bound41}
\ln|\Pii_2(\Xii)|=\fii_{2}(\Xii).
\end{equation}

Now, define
\begin{eqnarray}
\tvarphii_2(\bX)&\triangleq& \tfi_{2}(\bX)+\rho||\bx-\chii||^2\label{bound42}\\
&=&\sum_{\nu\in\clN_u}\lambdai_{\nu}\ai_{2,\nu}-2(\bi_2+\rho\chii)^T{\sf vec}(\bX)\nonumber\\
&&+{\sf vec}(\bX)^T (\tclDi_2+\rho I_{KM^2}){\sf vec}(\bX)\nonumber\\
&&+\rho||\chii||^2,\label{tvarphi}
\end{eqnarray}
which by (\ref{bound40}) and (\ref{bound41}) satisfies
\begin{equation}\label{bound43}
\varphii_2(\bX)\leq \tvarphii_2(\bX)\ \forall\ \bX,
\end{equation}
and
\begin{equation}\label{bound44}
\varphii_2(\Xii)\leq \tvarphii_2(\Xii).
\end{equation}
The subsequent holographic beamformer $\Xio$ is then obtained as the optimal solution of the following convex quadratic problem:
\begin{equation}\label{gsm7}
\max_{{\sf vec}(\bX)} \tvarphii_2(\bX)\quad\mbox{s.t.}
\quad (\ref{vec6}),
\end{equation}
which admits the closed-form solution
\begin{align}\label{gsm8}
&{\sf vec}(\Xio)\nonumber\\
&=\begin{cases}\begin{array}{l}\!\!\!\!(\tclDi_2+\rho I_{KM^2})^{-1}(\bi_2+\rho\chii)\\
\!\!\!\!\mbox{if}\ ||\sqrt{\tclDi_1}(\tclDi_2+\rho I_{KM^2})^{-1}(\bi_2+\rho\chii)||^2\leq P,\cr
\!\!\!\!(\tclDi_2+\rho I_{KM^2}+\tau \tclDi_1)^{-1}(\bi_2+\rho\chii)\\
\!\!\!\!\mbox{otherwise},
\end{array}
\end{cases}
\end{align}
where $\tau>0$ is found by bisection so that
\begin{equation}\label{gsm9}
||\sqrt{\tclDi_1}(\tclDi_2+\rho I_{KM^2}+\tau \tclDi_1)^{-1}(\bi_2+\rho\chii)||^2=P.
\end{equation}
The computational complexity of expression  (\ref{gsm8}) is on the order of ${\cal O}(M^2K)$.

Similarly to (\ref{bound5})-(\ref{bound7}), we have (\ref{sxal1})/(\ref{sxal1a}) as a consequence of (\ref{bound43}) and (\ref{bound44}) and $\tvarphii_2(\Xio)<\tvarphii_2(\Xii)$ whenever $\tvarphii_2(\Xio)\neq\tvarphii_2(\Xii)$, since $\Xio$ and $\Xii$ are the optimal solution and a feasible point of the minimization problem (\ref{ghb4}), respectively.
\subsection{Amplitude optimization and control}
The amplitude optimization and control aims for generating $\chio$ so that
\begin{align}
&\varphi(\Wio,\Xio,\chio)<\varphi(\Wio,\Xio,\chii)\nonumber\\
\Leftrightarrow&||\xio-\chiio||^2< ||\xio-\chii||^2\label{scal1}
\end{align}
for $\xio\triangleq {\sf vec}(\Xio)$. As such, $\chio$ is obtained by the closed-form (\ref{mgm1}).
\subsection{The algorithm and its convergence}
Algorithm \ref{alg3} provides the pseudo code for solving the problem (\ref{smm1}).
It follows from (\ref{swal1}), (\ref{sxal1}) and (\ref{scal1}) that
\begin{equation}\label{wxc1}
\varphi(\Wio,\Xio,\chio)< \varphi(\Wi,\Xii,\chi),
\end{equation}
as far as we have $\varphi(\Wio,\Xio,\chio)\neq \varphi(\Wi,\Xii,\chii)$, so the sequence
$\{(\Wi,\Xii,\chii)\}$ converges to $(\bar{W},\bar{X},\bar{\chi})$. Furthermore, by choosing a sufficiently large $\rho>0$, we can ensure that  $||{\sf vec}(\Xii)-\chii||^2\rightarrow 0$ as $\iota\rightarrow\infty$, which means that
$(\bar{W},\bar{X})$ is a feasible point for   (\ref{osm}), which turns out to be at least a local solution \cite{Betal06}.
The computational complexity of each iteration for solving the problem (\ref{smm1}) is equivalent to that of Algorithm \ref{alg2}.
\begin{algorithm}[!t]
	\caption{Scalable-complexity algorithm for soft max-min optimization} \label{alg3}
	\begin{algorithmic}[1]
	\State \textbf{Initialization:} Initialize a feasible point $(X^{(0)},W^{(0)},\chi^{(0)})$  for (\ref{smm1}). Set $\iota=0$.
		\State \textbf{Repeat until convergence:} Generate  $\Wio$ by the closed-form
 (\ref{bsm11}), $\Xio$  by the closed-form  (\ref{gsm8}), and $\chio$ by the closed form (\ref{mgm1}). Reset $\iota \leftarrow \iota+1$.
 \State \textbf{Output} $(W^{opt},X^{opt})=(\Wi,\Xii)$.
	\end{algorithmic}
\end{algorithm}
\section{Numerical Examples}
In this section, a multi-user system having a BS equipped with a RHS relying on $12\times 12$ radiation elements for supporting $4$ UEs is considered. The UEs are randomly distributed within a cell radius of $150$ meters. The height of the BS and the UEs are set to $10$ meters and $1.5$ meters, respectively. The carrier frequency is set to $28$ GHz, with a system bandwidth of $100$ MHz. The noise power density is set to $-174$ dBm/Hz. The propagation vector on the RHS is set to $|\kappa_s|=\sqrt{3}|\kappa|$, where $\sqrt{3}$ is refractive index of the substrate of the waveguide \cite{Denetal22twc}. Adopting the channel structure from \cite{Denetal22twc}, the propagation environment between the BS and UE ${\nu}$ is characterized by a millimeter wave channel $H_{\nu} = \sqrt{10^{-\beta_{\nu}/10}}\sqrt{\frac{M^2D}{L}}\sum_{\ell=1}^{L} \alpha_{\nu,\ell} a_r\left(\phi_{\nu,\ell}^r\right) a_t^H \left(\phi_{\nu,\ell}^t,\theta_{\nu,\ell}^t \right)$,
where the path-loss of UE ${\nu}$ at distance $d_{\nu}$ is set to $\beta_{\nu}=53.22+35.3\log 10(d_{\nu})$ (in dB) \cite{Rappa17,3GPPTR38}, the number of paths is set to $L = 15$ \cite{SY16}, and the complex gain of the $\ell$-th path is set to $\alpha_{\nu,\ell} \sim \mathcal{CN}(0,1)$. Furthermore, $(\phi_{\nu,\ell}^t, \theta_{\nu,\ell}^t)$ and $\phi_{\nu,\ell}^r$ represent the uniformly random angle of departure (AoD) and angle of arrival (AoA) of $\phi_{\nu,\ell}^t, \phi_{\nu,\ell}^r\in [0,2\pi)$ and $\theta_{\nu,\ell}^t \in [-\pi/2,\pi/2)$, respectively. Let $\lambda$ denote the wavelength, and assuming that the element spacing $d_s$ of the RHS along both the $x$-axis and $y$-axis is set to quarter of the wavelength, and the antenna spacing $d_u$ at the UE side is half-wavelength, the normalized transmit and receive antenna array response vectors $a_t \left(\phi_{\nu,\ell}^t,\theta_{\nu,\ell}^t \right)$ and $a_r\left(\phi_{\nu,\ell}^r\right)$ are formulated as
\begin{equation}
\begin{array}{c}
	\begin{aligned}
	&a_t \left(\phi_{\nu,\ell}^t,\theta_{\nu,\ell}^t \right)\\
	=&\frac{1}{\sqrt{M^2}} \left[1, e^{j \frac{2\pi}{\lambda}[d_s\cos(\phi_{\nu,\ell}^t)\sin(\theta_{\nu,\ell}^t) + d_s\sin(\phi_{\nu,\ell}^t )\sin(\theta_{\nu,\ell}^t)] },\hdots,\right.\\
	&\left. e^{j \frac{2\pi}{\lambda}[(M-1)d_s\cos(\phi_{\nu,\ell}^t )\sin(\theta_{\nu,\ell}^t) + (M-1)d_s\sin(\phi_{\nu,\ell}^t )\sin(\theta_{\nu,\ell}^t)]} \right]^T,
	\end{aligned}
\end{array}
\end{equation}
and
\begin{align}
	&a_r \left(\phi_{\nu,\ell}^r \right)\nonumber\\
	=&\frac{1}{\sqrt{D}} \left[ 1,e^{j\frac{2\pi}{\lambda}d_u \sin(\phi_{\nu,\ell}^r)} ,
	\cdots, e^{j\frac{2\pi}{\lambda}d_u (D-1)\sin(\phi_{\nu,\ell}^r)}\right]^T.
\end{align}

Upon exploiting the superposition property of the holographic pattern, which maps all transmitted data onto a single holographic pattern on the RHS, the initial holographic beamformers are computed as the arithmetic mean of the normalized radiation amplitude corresponding to each object beam, i.e., $\mu_{k,m,m'}=\frac{1}{N_u} \sum_{\nu}^{N_u} \frac{\Re\{\psi_{k,m,m'}(d_{m,m'}^k,\theta_{\nu},\varphi_{\nu})\}+1}{2}$.

The following legends are used to specify the proposed implementations: $(i)$ {\tt MM} denotes the convex-quadratic-solver-based max-min rate Algorithm \ref{alg1}; $(ii)$ {\tt SR} denotes the scalable-complexity sum-rate Algorithm \ref{alg2}; $(iii)$ {\tt SMM} denotes the scalable-complexity soft max-min rate Algorithm \ref{alg3}.

\subsection{Algorithm validation in a simple single-antenna UE scenario}
Table \ref{table:complexity} illustrates the computational complexity of the proposed algorithms. The {\tt MM} algorithm exhibits cubic complexity, whereas the {\tt SR} and {\tt SMM} algorithms demonstrate scalable complexity. Due to the high computational complexity of the convex-solver-based {\tt MM} algorithm, we commence by examining the single-antenna UE scenario, where the BS is equipped with $K = 4$ feeds. For the {\tt MM} algorithm, we initialize the holographic beamformer $\bX^{(0)}$ by randomly generating elements $\bx_{k,m,m'}^{(0)},(k,m,m')\in\clK\times\clM\times\clM$ that satisfy the amplitude constraint (\ref{holo10c}). Similarly, we randomly generate the baseband beamformer $\bW^{(0)}$ satisfying the power constraint (\ref{holo10b}). For the {\tt SR} and {\tt SMM} algorithms, we utilize the same $\bX^{(0)}$ and $\bW^{(0)}$ generated for the {\tt MM} algorithm. Finally, the auxiliary variable $\bchi_{k,m,m'}^{(0)},(k,m,m')\in\clK\times\clM\times\clM$ is initialized by adding a random perturbation to  $\bx_{k,m,m'}^{(0)}$ while ensuring the amplitude constraint (\ref{gm1c}) is satisfied.

\begin{table}[t]
	\centering
	\caption{Complexity of proposed algorithms}
	\begin{tabular}{lccc}
		\toprule
		& {\tt MM} Alg. 1  & {\tt SR} Alg. 2  & {\tt SMM} Alg. 3 \\
		\midrule
		Baseband iter.    & ${\cal O}(N_u^3K^3D^3)$ & ${\cal O}(N_uKD)$ & ${\cal O}(N_uKD)$ \\
		Holographic iter. & ${\cal O}((M^2K)^4)$    & ${\cal O}(M^2K)$  & ${\cal O}(M^2K)$  \\
		Amplitude iter.   &                         & ${\cal O}(M^2K)$  & ${\cal O}(M^2K)$  \\
		\bottomrule
	\end{tabular}
	\label{table:complexity}
\end{table}

\begin{figure}[t]
	\centering
	\begin{subfigure}[c]{0.4\textwidth}
		\centering
		\includegraphics[width=6.2cm]{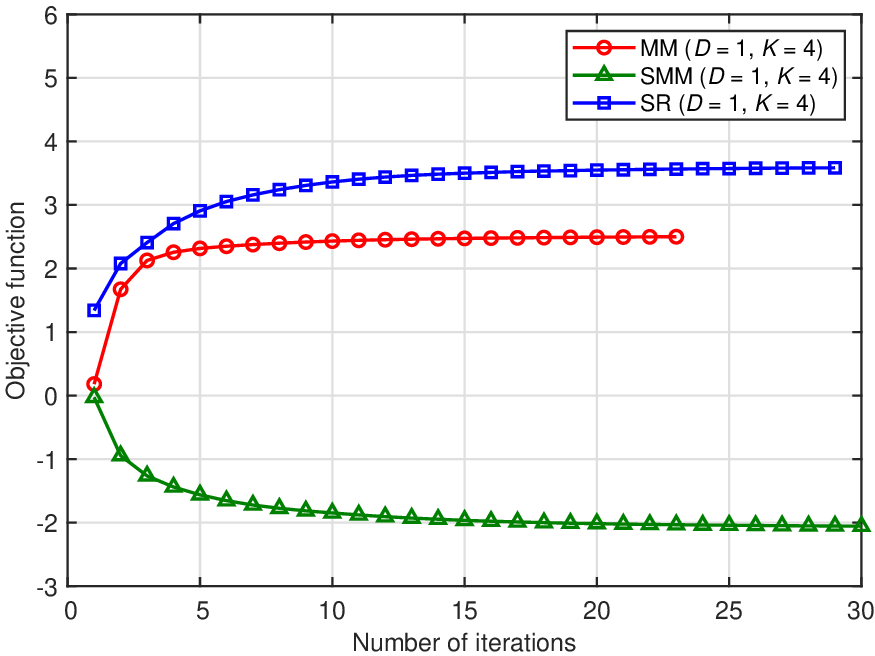}
		\caption{Objective function vs. number of iterations}
		\label{fig:conv_obj_fun_UE4_Feed4}
	\end{subfigure}
	\hfill
	\begin{subfigure}[c]{0.4\textwidth}
		\centering
		\includegraphics[width=6.2cm]{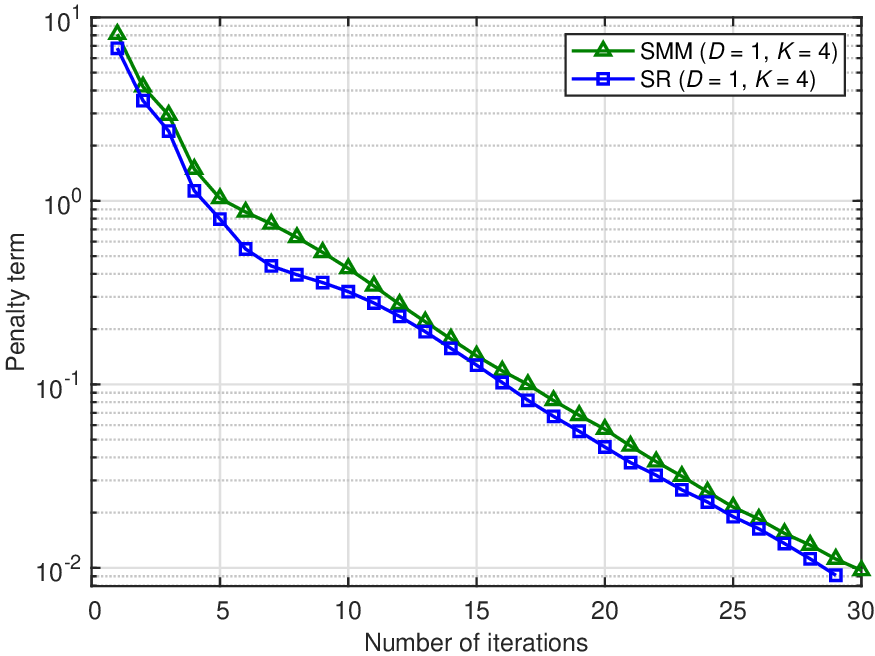}
		\caption{Penalty terms vs. number of iterations}
		\label{fig:conv_penalty_UE4_Feed4}
	\end{subfigure}
	\hfill
	\begin{subfigure}[c]{0.4\textwidth}
		\centering
		\includegraphics[width=6.2cm]{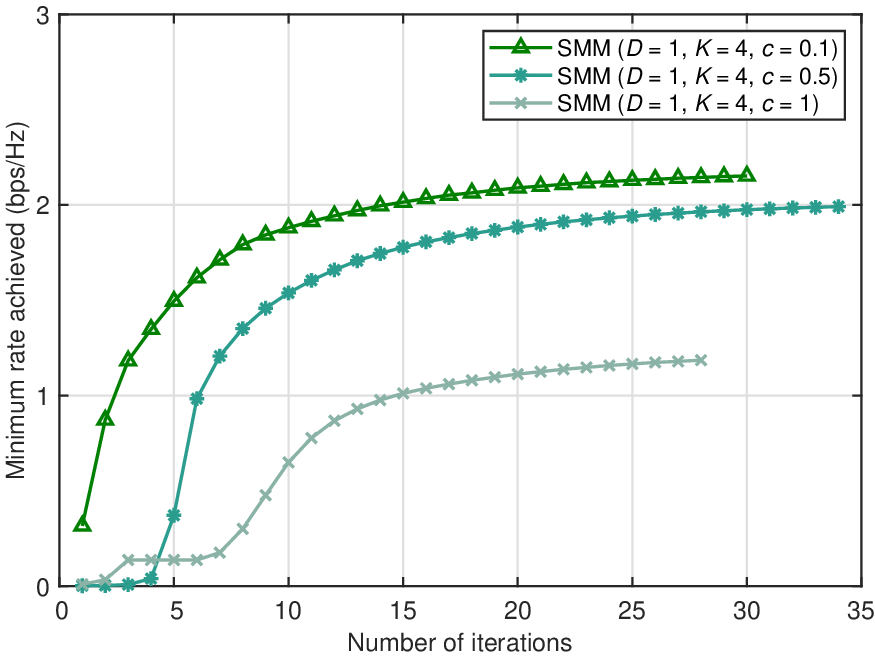}
		\caption{Minimum rate achieved by the {\tt SMM} algorithm vs. number of iterations}
		\label{fig:conv_min_rate_UE4_Feed4_SMM}
	\end{subfigure}
	\caption{Convergence performance of proposed algorithms for $D = 1$ at $P = 24$ dBm.}
	\label{fig:conv_UE4_Feed4}
\end{figure}

Fig. \ref{fig:conv_UE4_Feed4} characterizes the convergence performance of the proposed algorithms for $D = 1$ at a transmit power of $P = 24$ dBm. We use the arithmetic mean value of the sum-rate objective to offer a compact representation of the convergence patterns of all the proposed algorithms. For the convex-solver-based {\tt MM} algorithm, we consider the algorithm to be converged, when the error tolerance of the objective function reaches $10^{-3}$. For both the {\tt SMM} and {\tt SR} algorithms, a penalized optimization approach is used for addressing the real-domain amplitude constraints in holographic beamforming. The algorithm terminates when the penalty term drops below $10^{-2}$.
To find the optimal solution at satisfactory convergence speed, we commence with a penalty factor $\rho$ that aligns the magnitude of the penalty term with that of the objective. Specifically, we set $\rho=f_S(\bW^{(0)},\bX^{(0)})/||{\sf vec}(\bX^{(0)})-\bchi^{(0)}||^2$ for the {\tt SR} algorithm, and set $\rho=\ln|\Pi(\bW^{(0)},\bX^{(0)})|/||{\sf vec}(\bX^{(0)})-\bchi^{(0)}||^2$ for the {\tt SMM} algorithm. The value of $\rho$ is then slightly increased. During our simulations, we observed that updating $\rho\rightarrow 1.2\rho$ whenever $||x^{(\iota+1)}-\chi^{(\iota+1)}||^2>0.9||x^{(\iota)}-\chi^{(\iota)}||^2$ not only results in a gradual reduction of the penalty term towards zero but also achieves favorable user rates. Note that this updating strategy is flexible and can be adjusted to balance the convergence speed vs. rate for specific scenarios. Additionally, the impact of the coefficient $c$ on the minimum rate of the {\tt SMM} algorithm is evaluated by analyzing its convergence behavior for different values of $c$, as illustrated in Fig. \ref{fig:conv_min_rate_UE4_Feed4_SMM}. The selection of $c$ will be further discussed below.

\begin{table*}[t]
	\centering
	\caption{The minimum rate achieved vs. $c$ under different transmit powers $P$ using the {\tt SMM} algorithm for $D = 1$.}
	\begin{tabular}{lccccc}
		\toprule
		& $P = 16$ dBm & $P = 18$ dBm & $P = 20$ dBm & $P = 22$ dBm & $P = 24$ dBm \\
		\midrule
		$c = 1$   & 0.00 & 0.01 & 0.01 & 0.55 & 1.13 \\
		$c = 0.5$ & 0.06 & 0.26 & 0.83 & 1.56 & 2.08 \\
		$c = 0.1$ & 0.68 & 0.94 & 1.28 & 1.67 & 2.10 \\
		\bottomrule
	\end{tabular}
	\label{table:min_rate_vs_c_UE4_Feed4}
\end{table*}

\begin{figure}[t]
	\centering
	\includegraphics[width=6.6cm]{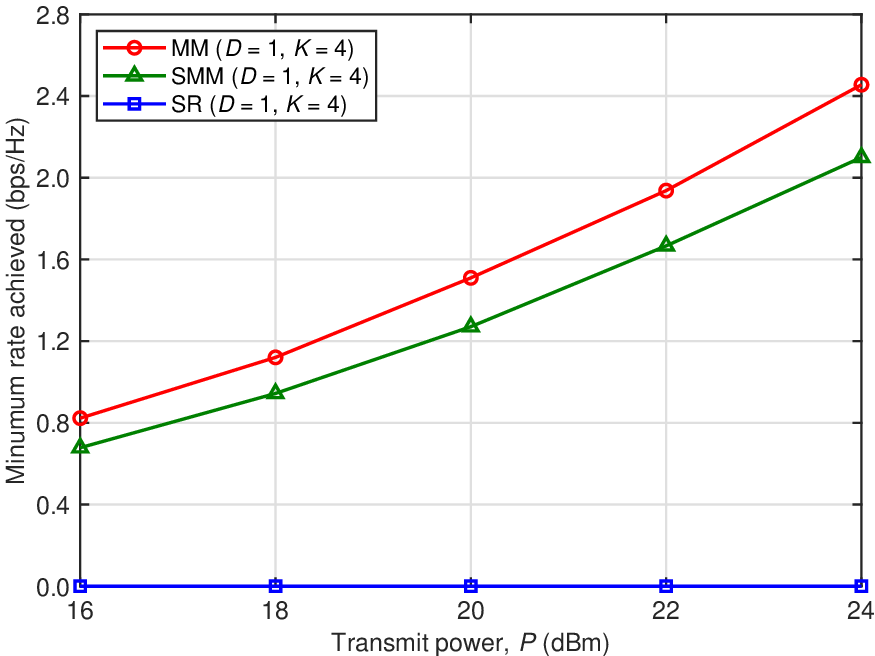}
	\caption{The minimum rate vs. transmit power $P$ for $D = 1$.}
	\label{fig:min_rate_vs_P_UE4_Feed4}
\end{figure}

\begin{figure}[t]
	\centering
	\includegraphics[width=6.6cm]{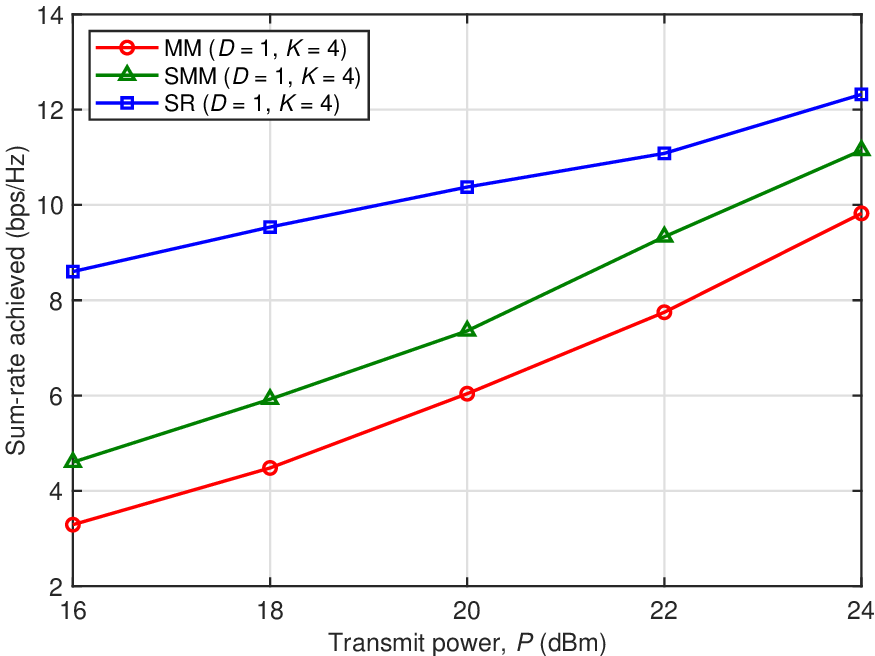}
	\caption{The sum-rate vs. transmit power $P$ for $D = 1$.}
	\label{fig:sum_rate_vs_P_UE4_Feed4}
\end{figure}

\begin{figure}[t]
	\centering
	\includegraphics[width=6.6cm]{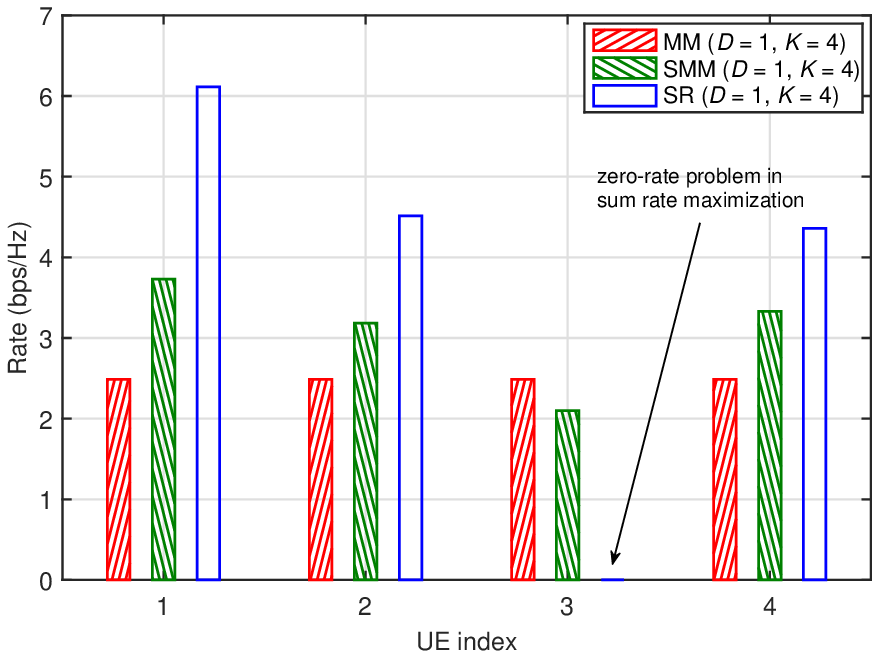}
	\caption{The rate distributions for $D = 1$ at $P = 24$ dBm.}
	\label{fig:rate_distrib_UE4_Feed4_P24}
\end{figure}

Fig. \ref{fig:min_rate_vs_P_UE4_Feed4} illustrates the minimum rate achieved by the proposed algorithms. As expected, the {\tt MM} algorithm yields the highest minimum rate. However, the {\tt SR} algorithm results in zero-rate, making it unsuitable for multi-user communication. The {\tt SMM} algorithm, which maximizes an approximation of the minimum rate objective, provides a solution, which is approximately 85\% of the minimum rate obtained by maximizing the original minimum rate objective. Moreover, the {\tt SMM} algorithm offers an appealing scalable complexity compared to the cubic complexity of the {\tt MM} algorithm.

The sum-rate performance attained by the proposed algorithms is depicted in Fig. \ref{fig:sum_rate_vs_P_UE4_Feed4}. The {\tt SMM} algorithm outperforms the {\tt MM} algorithm in terms of its sum-rate. Additionally, as the transmit powers increase, the sum-rate achieved by the {\tt SMM} algorithm is approaching the value obtained by directly maximizing the sum-rate.

It is noteworthy that the coefficient $c$ in the {\tt SMM} algorithm will influence the minimum rate achieved, thus requires careful selection. Table \ref{table:min_rate_vs_c_UE4_Feed4} provides guidance for selecting $c$ by comparing the minimum rate achieved under various values of $c$, with $c = 0.1$ yielding the highest minimum rate.

Fig. \ref{fig:rate_distrib_UE4_Feed4_P24} shows the distribution of individual rates achieved by the proposed algorithms, illustrating the zero-rate problem inherent in sum-rate maximization. Observe that the max-min strategy excessively prioritizes fairness at the expense of users with favorable propagation conditions, while maximizing the sum-rate results in allocating a large portion of the total rate to a few UEs having favorable channel conditions, leading to zero-rate allocation for certain UEs. By contrast, the {\tt SMM} algorithm yields a more equitable rate distribution without significantly compromising the total rate, while achieving a total rate higher than that of the {\tt MM} algorithm.

Note that given the parameter settings used in this subsection, the holographic beamforming optimization stage involves $M\times M\times K = 576$ optimization variables. As the size of the RHS $M\times M$ or the number of feeds $K$ continue to expand, it becomes challenging for the {\tt MM} algorithm to strike a balance between the rate attained and the computational efficiency, since this algorithm optimizes the minimum rate using a convex solver having cubic complexity. Therefore, we exclude the {\tt MM} algorithm in the subsequent multi-antenna UE scenarios.

\begin{table*}[t]
	\centering
	\caption{The average number of zero-rate UEs in maximizing the sum-rate for $D = 1\ \&\ 2$.}
	\begin{tabular}{lccccc}
		\toprule
		& $P = 16$ dBm & $P = 18$ dBm & $P = 20$ dBm & $P = 22$ dBm & $P = 24$ dBm \\
		\midrule
		{\tt SR} $(D = 1)$ & 1.6 & 1.4 & 1.1 & 0.8 & 0.7 \\
		{\tt SR} $(D = 2)$ & 1.6 & 1.3 & 1.2 & 0.9 & 0.8\\
		\bottomrule
	\end{tabular}
	\label{table:zero_rate_vs_P_UE4_Feed8}
\end{table*}

\begin{figure}[t]
	\centering
	\begin{subfigure}[c]{0.4\textwidth}
		\centering
		\includegraphics[width=6.6cm]{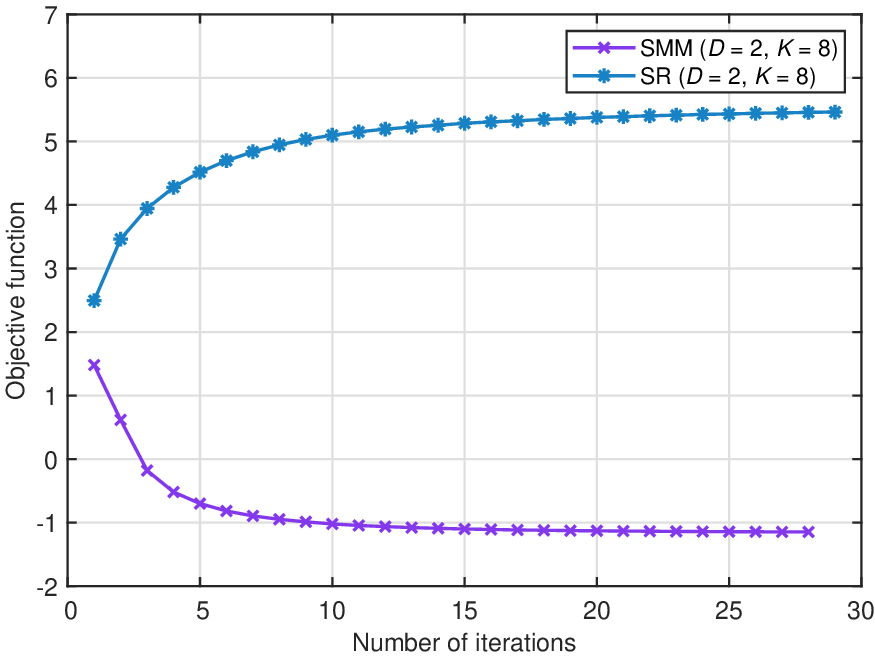}
		\caption{Objective function vs. number of iterations}
		\label{fig:conv_obj_fun_UE4_Feed8}
	\end{subfigure}
	\hfill
	\begin{subfigure}[c]{0.4\textwidth}
		\centering
		\includegraphics[width=6.6cm]{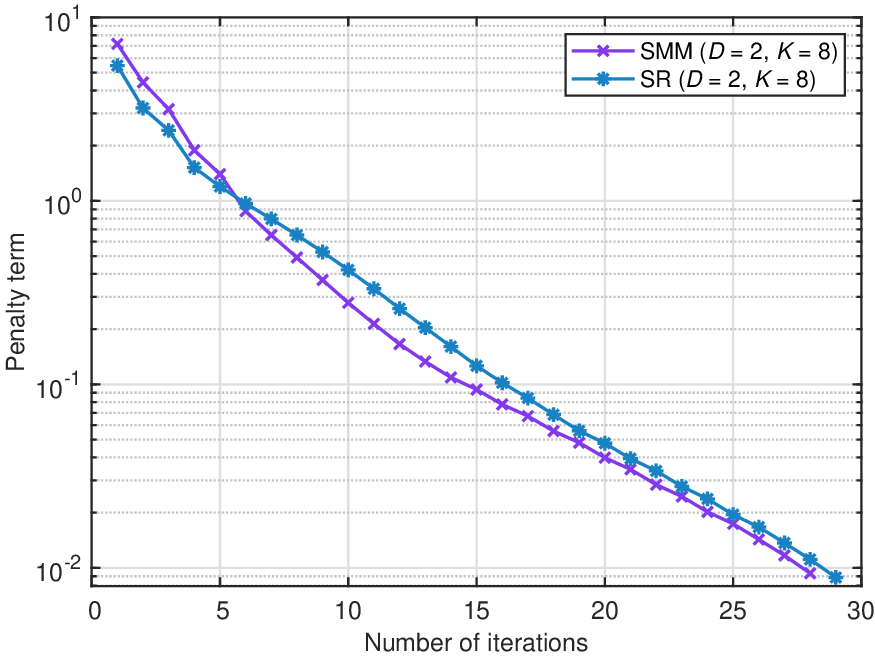}
		\caption{Penalty terms vs. number of iterations}
		\label{fig:conv_penalty_UE4_Feed8}
	\end{subfigure}
	\caption{Convergence performance of proposed algorithms for $D = 2$ at $P = 24$ dBm.}
	\label{fig:conv_UE4_Feed8}
\end{figure}

\subsection{Comparison of single-antenna and multi-antenna UE scenarios}
Next, we examine the scenario of multi-antenna UEs having $D = 2$ for comparing it to the single-antenna UE scenario. To achieve a fair assessment, we assume that the BS is equipped with $K = 8$ feeds, ensuring that the number of feeds exceeds or equals the total number of UE antennas. Note that in the multi-antenna UE scenario having the aforementioned parameters, the holographic beamforming optimization stage involves $M\times M\times K = 1152$ optimization variables, ruling out the use of convex-solver-based algorithms for practical implementations.

\begin{figure}[!t]
	\centering
	\includegraphics[width=6.6cm]{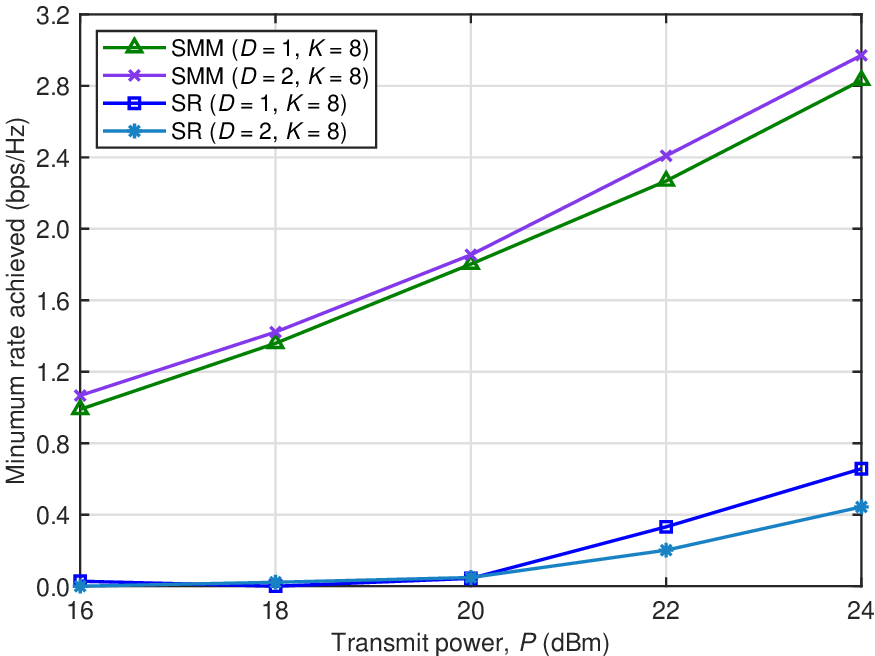}
	\caption{The minimum rate vs. transmit power $P$ for $D = 1\ \&\ 2$.}
	\label{fig:min_rate_vs_P_UE4_Feed8}
\end{figure}

\begin{figure}[!t]
	\centering
	\includegraphics[width=6.6cm]{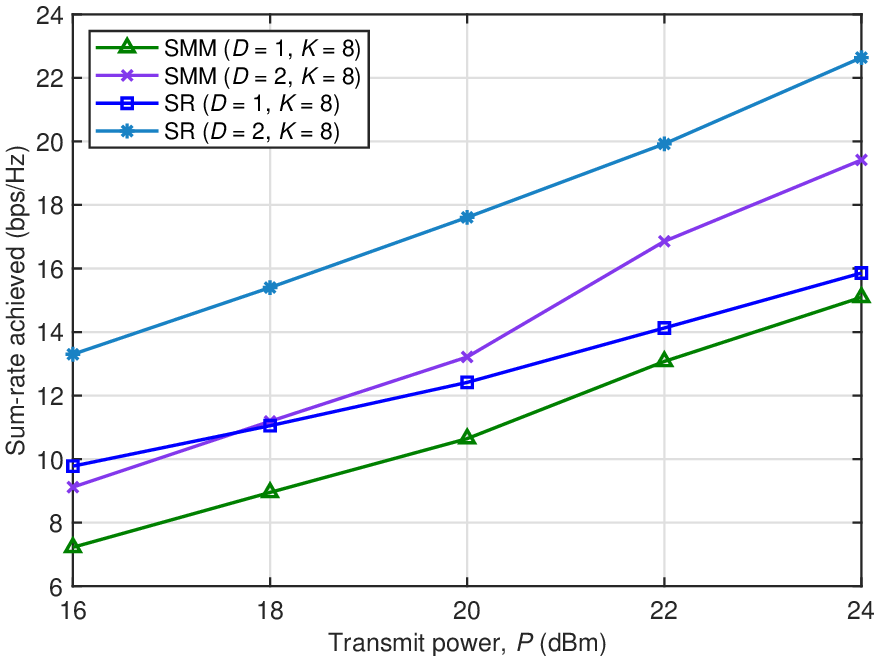}
	\caption{The sum-rate vs. transmit power $P$ for $D = 1\ \&\ 2$.}
	\label{fig:sum_rate_vs_P_UE4_Feed8}
\end{figure}

We initially characterize the convergence performance of the {\tt SMM}\footnote{The {\tt SMM} algorithm with $c = 0.5$ exhibits the best minimum rate performance for $D = 2$ with a transmit power of $P = 24$ dBm.} and {\tt SR} algorithms at $P = 24$ dBm in the multi-antenna UE scenario, where the rate function is formulated as a log determinant. Fig. \ref{fig:conv_UE4_Feed8} demonstrates the efficiency of our approaches in addressing the optimization of nonlinear log determinant functions.

Fig. \ref{fig:min_rate_vs_P_UE4_Feed8} compares the minimum rates achieved by the scalable-complexity {\tt SMM} and {\tt SR} algorithms in the single-antenna and multi-antenna UE scenarios. It emphasizes the inadequacy of sum-rate optimization for multi-user communication, even upon increasing the number of UE antennas. We present Table \ref{table:zero_rate_vs_P_UE4_Feed8} in support of this observation. The results of both Fig. \ref{fig:min_rate_vs_P_UE4_Feed8}  and Table \ref{table:zero_rate_vs_P_UE4_Feed8}  reveal that while maximizing the sum-rate for increased transmit power reduces the probability of allocating zero-rate to certain UEs, it does not guarantee reliable connections for all UEs in all possible scenarios.

Fig. \ref{fig:sum_rate_vs_P_UE4_Feed8} compares the sum-rate obtained through the maximization of sum-rate and soft minimum rate. For $D = 2$, there is a significant sum-rate enhancement compared to $D = 1$. The {\tt SMM} algorithm is capable of generating a sum-rate that closely approaches the one achieved through direct maximization. Moreover, with the number of feeds increased to 8, the {\tt SMM} algorithm yields approximately 95\% of the sum-rate achieved by the {\tt SR} algorithm in the single-antenna UE scenario. In the case of two-antenna UEs, the {\tt SMM} algorithm still achieves approximately 85\% of the sum-rate. Thus, we may conclude that the {\tt SMM} solutions strike an attractive balance in achieving both an appealing minimum rate and sum-rate, without significantly sacrificing one objective against another, as it is often the case in conventional stand-alone minimum rate or sum-rate maximization problems.

\section{Summary and Conclusions}
The downlink of a network was considered, where the base station is equipped with holographic reconfigurable surfaces transmitting multi-stream information to multiple users having multi-antenna arrays.
To ensure rate-fairness, we explored the joint design of holographic and baseband beamformers to enhance the users' minimum rates. Specifically, we have developed an algorithm that utilizes quadratic solvers for improving the minimum rates of users.
Furthermore, we introduced a new surrogate optimization problem aimed at achieving both a minimum and a sum-rate comparable to that attained by max-min rate optimization and sum-rate maximization. This surrogate optimization problem has demonstrated significant computational benefits, since its computation relies on a scalable algorithm that iterates by evaluating closed-form expressions.

\section*{Appendix A: Fundamental Tight Minorant and Majorant}
According to \cite[p. 366]{Tuybook},
a function $\bar{f}$ is said to be a tight minorant of a function $f$ over the domain $\mbox{dom}(f)$ at
a point $\bar{z}\in\mbox{dom}(f)$ if it serves as a minorant of $f$:
\begin{equation}\label{dom1}
	f(\bz)\geq \bar{f}(\bz)\ \forall\ \bz\in\mbox{dom}(f),
\end{equation}
and matches $f$ at $\bar{z}$:
\begin{equation}\label{dom2}
	f(\bar{z})=\bar{f}(\bar{z}).
\end{equation}
Note that $f(z_{\max})\geq f(\bar{z})$ for $z_{\max}\triangleq \mbox{arg}\max_{\bz\in\mbox{dom}(f)}\bar{f}(\bz)$. In other words,
maximizing a tight minorant of $f$ helps identify  a better
point than $\bar{z}$ for the maximization problem considered.

Similarly, $\bar{f}$ is a tight majorant of $f$ over $\mbox{dom}(f)$ at
$\bar{z}\in\mbox{dom}(f)$ if it serves as a majorant of $f$:
\begin{equation}\label{dom1a}
	f(\bz)\leq \bar{f}(\bz)\ \forall\ \bz\in\mbox{dom}(f),
\end{equation}
and matches $f$ at $\bar{z}$ (see (\ref{dom2})). Since we have $f(z_{\min})\leq f(\bar{z})$ for $z_{\min}\triangleq \mbox{arg}\min_{\bz\in\in\mbox{dom}(f)}\bar{f}(\bz)$,
minimizing a tight majorant of $f$ helps identify  a better
point than $\bar{z}$ for the minimization problem.

The following matrix inequality holds for all $\bV$, $\bar{V}$, and positive definite $\bY$ and $\bar{Y}$ of appropriate dimension,  as established in \cite{TTN16}:
\begin{align}
	\ln|\mI+[\bV]^2\bY^{-1}|\geq& \ln|\mI+[\bar{V}]^2\bar{Y}^{-1}|-\la[\bar{V}]^2\bar{Y}^{-1}\ra\nonumber\\
	& +2\Re\{\la \bar{V}^H\bar{Y}^{-1}\bV\ra\}\nonumber\\
	&-\la \bar{Y}^{-1}-\left([\bar{V}]^2+\bar{Y}\right)^{-1}, [\bV]^2+\bY\ra.\label{fund5}
\end{align}
Considering both sides of (\ref{fund5}) as functions of $(\bV,\bY)$, the right hand side equals the left hand side at the point $(\bar{V},\bar{Y})$. Consequently,
the right hand side  provides a tight minorant of the left hand side at $(\bar{V},\bar{Y})$. Maximizing the right hand side of (\ref{fund5}), referred to as tight minorant maximization, produces an improved point over $(\bar{V},\bar{Y})$ for the sake of maximizing the left hand side of (\ref{fund5}).

For $\bV\triangleq (\bV_1,\dots, \bV_{N_u})$ and $\bY\triangleq (\bY_1,\dots, \bY_{N_u})$ associated
with matrices $\bV_{\nu}$ and Hermitian symmetric matrices $\bY_{\nu}$, $\nu=1,\dots, N_u$
of appropriate dimension,  and for
\begin{equation}\label{ap31a}
	\Pi(\bV,\bY)\triangleq \sum_{\nu=1}^{N_u}(\mI-\bV^H_{\nu}\bY^{-1}_{\nu}\bV_{\nu}),
\end{equation}
in the domain:
\begin{eqnarray}\label{ap32}
	\mbox{dom}(\Pi)\triangleq
	\left\{(\bV,\bY): [\bV_{\nu}]^2\prec \bY_{\nu}, \nu=1,\dots, N_u\right\},
\end{eqnarray}
the following inequality holds for all $(\bV,\bY)$ and $(\bar{V},\bar{Y})$, as established in \cite{Tuaetal24}:
\begin{align}
	\ln\Big|\Pi(\bV,\bY)\Big|\leq&\ln\Big|\Pi(\bar{V},\bar{Y})\Big|+\sum_{\nu=1}^{N_u}\la \Pi^{-1}(\bar{V},\bar{Y})\bar{V}^H_{\nu}\bar{Y}_{\nu}^{-1}\bar{V}_{\nu}\ra\nonumber\\
	&-2\sum_{\nu=1}^{N_u}\Re\{\la \Pi^{-1}(\bar{V},\bar{Y}) \bar{V}^H_{\nu}\bar{Y}_{\nu}^{-1}\bV_{\nu}\ra\}\nonumber\\
	&+\sum_{\nu=1}^{N_u}\la \bar{Y}_{\nu}^{-1}\bar{V}_{\nu} \Pi^{-1}(\bar{V},\bar{Y})\bar{V}^H_{\nu}\bar{Y}_{\nu}^{-1}\bY_{\nu}\ra.\label{ap6}
\end{align}
Considering both sides of (\ref{ap6}) as functions of $(\bV,\bY)$,
the right hand side provides a tight majorant of the left hand side at $(\bar{V},\bar{Y})$, since these two coincide at this point. Analogously, minimizing the right hand side of (\ref{ap6}), referred to as tight majorant minimization, produces an improved point over $(\bar{V},\bar{Y})$ for the sake of minimizing the left hand side of (\ref{ap6}).
\section*{Appendix B: derivation of (\ref{vec5})}
With the tight minorant $\tri_{2,\nu}(\bX)$ of $r_{2,\nu}(\bX)$ as
defined in (\ref{hb3})-(\ref{hb4}), we have
\begin{eqnarray}
\sum_{\nu\in\clN_u}\tri_{2,\nu}(\bX)&=&\sum_{\nu\in\clN_u}\left[\ai_{2,\nu}+2\Re\{\la \clBi_{2,\nu}\bX\ra\}\right.\nonumber\\
&&\left.-\la \tclCi_{2,\nu},
\bX\clAio\bX^T\ra\right]\nonumber\\
&=&\sum_{\nu\in\clN_u}\ai_{2,\nu}
+2\Re\{\la\tclBi_2\bX\ra\}\nonumber\\
&&-\la\tclCi_2,\bX\clAio\bX^T\ra,\label{ghb2}
\end{eqnarray}
where $(\tclBi_2,\tclCi_2)$ and $\clAio$ are defined in (\ref{ghb3}) and (\ref{bb2b}), respectively, with $(\ai_{2,\nu},\clBi_{2,\nu},\tclCi_{2,\nu})$  defined in (\ref{hb4}).

Furthermore,
\begin{equation}\label{vec1}
\Re\{\la\tclBi_2\bX\ra\}= (\bi_2)^T{\sf vec}(\bX)
\end{equation}
for $\bi_2$ defined in (\ref{vec2}), and\footnote{The intermediate step follows from the vectorization identity ${\sf vec}(AXB)=(B^T\otimes A){\sf vec}(X)$, where $A$, $X$, and $B$ are matrices of appropriate dimensions.}
\begin{align}
\la\tclCi_2,\bX\clAio\bX^T\ra
=&||{\sf vec}\left(\sqrt{\tclCi_2}\bX\sqrt{\clAio}\right)||^2\nonumber\\
=&{\sf vec}^T(\bX)\left((\clAio)^T\otimes \tclCi_2 \right){\sf vec}(\bX)\nonumber\\
=&{\sf vec}^T(\bX) \tclDi_2{\sf vec}(\bX),\label{vec3}
\end{align}
for $\tclDi_2$ defined in (\ref{vec4}).

Finally, (\ref{vec5}) follows by substituting the expressions for $\Re\{\la\tclBi_2,\bX\ra\}$ from (\ref{vec1}) and for $\la\tclCi_2,\bX\clAio\bX^T\ra$ from (\ref{vec3}) into (\ref{ghb2}).
\section*{Appendix C: derivation of (\ref{gvec5})}
We have
\begin{eqnarray}
\mbox{RHS of}\ (\ref{gsm3})&=&\ai_2-2\Re\{\la \tclBi_2\bX\ra\}\nonumber\\
&&+\sum_{\nu\in\clN_u}\la\tclCi_{2,\nu},\bX\clAio_{\nu}\bX^T\ra.\label{gsm4}
\end{eqnarray}
with $\ai_2$ and $(\tclBi_2,\tclCi_{2,\nu}, \clAio_{\nu})$ defined in (\ref{gsm5a}) and
(\ref{gsm6}), respectively.

Furthermore,
\begin{equation}\label{gvec1}
\Re\{\la\tclBi_2\bX\ra\}= (\bi_2)^T{\sf vec}(\bX)
\end{equation}
with $\bi_2$ defined in (\ref{gvec2}),  and
\begin{align}
\la\tclCi_{2,\nu},\bX\clAio_{\nu}\bX^T\ra
=&||{\sf vec}\left(\sqrt{\tclCi_{2,\nu}}\bX\sqrt{\clAio_{\nu}}\right)||^2\nonumber\\
=&{\sf vec}(\bX)^T\left((\clAio_{\nu})^T\otimes \tclCi_{2,\nu} \right){\sf vec}(\bX)\nonumber\\
=&{\sf vec}(\bX)^T \tclDi_{2,\nu}{\sf vec}(\bX),\label{gvec3}
\end{align}
with $\tclDi_{2,\nu}$ defined in (\ref{gvec4}).

Finally, (\ref{gvec5}) follows by substituting the expressions for $\Re\{\la\tclBi_2\bX\ra\}$ from (\ref{gvec1}) and for $\la\tclCi_{2,\nu},\bX\clAio_{\nu}\bX^T\ra$ from (\ref{gvec3}) into (\ref{gsm4}).

\bibliographystyle{IEEEtran}
\balance \bibliography{mmwave}

\end{document}